\documentclass{optica-article}

\journal{opticajournal} 
\pdfoutput=1
\articletype{Research Article}

\usepackage{lineno,gensymb,amsmath,braket}

\begin{document}

\title{An accurate vector optically pumped magnetometer with microwave-driven Rabi frequency measurements}

\author{Christopher Kiehl,\authormark{1,2,*} Thanmay S. Menon,\authormark{1,2} Svenja Knappe,\authormark{3,4} Tobias Thiele\authormark{5} and Cindy A. Regal\authormark{1,2}}

\address{\authormark{1}JILA, National Institute of Standards and Technology and University of Colorado, Colorado 80309, USA\\
\authormark{2}Department of Physics, University of Colorado, Boulder, Colorado 80309, USA\\
\authormark{3}Paul M. Rady Department of Mechanical Engineering, University of Colorado, Boulder, Colorado 80309, USA\\
\authormark{4}FieldLine Inc., Boulder CO 80301, USA\\
\authormark{5}Zurich Instruments AG, CH-8005 Zurich, Switzerland}

\email{\authormark{*}christopher.kiehl@colorado.edu} 

\begin{abstract*} 
Robust calibration of vector optically pumped magnetometers (OPMs) is a nontrivial task, but increasingly important for applications requiring high-accuracy such as magnetic navigation, geophysics research, and space exploration. Here, we showcase a vector OPM that utilizes Rabi oscillations driven between the hyperfine manifolds of $^{87}$Rb to measure the direction of a DC magnetic field against the polarization ellipse structure of a microwave field. By relying solely on atomic measurements---free-induction decay (FID) signals and Rabi measurements across multiple atomic transitions---this sensor can detect drift in the microwave vector reference and compensate for systematic shifts caused by off-resonant driving, nonlinear Zeeman (NLZ) effects, and buffer gas collisions. To facilitate dead-zone-free operation, we also introduce a novel Rabi measurement that utilizes dressed-state resonances that appear during simultaneous Larmor precession and Rabi driving (SPaR). These measurements, performed within a microfabricated vapor cell platform, achieve an average vector accuracy of 0.46 mrad and vector sensitivities down to 11 $\mu$rad$/\sqrt{\text{Hz}}$ for geomagnetic field strengths near 50 $\mu$T. This performance surpasses the challenging 1-degree (17 mrad) accuracy threshold of several contemporary OPM methods utilizing atomic vapors with an electromagnetic vector reference.
\end{abstract*}

\section{Introduction}
Optically pumped magnetometers (OPMs) based on hot atomic vapors are increasingly important for both fundamental science and technological applications. These sensors utilize atomic spin precession, achieving record precision down to attotesla scales~\cite{kominis2003subfemtotesla,budker2007optical,dang2010ultrahigh,sheng2013subfemtotesla}. Because the precession frequency is only sensitive to the magnetic field strength with respect to well-known atomic constants, OPMs are highly accurate and less dependent on the sensor orientation compared to directional magnetometers, such as fluxgates or superconducting quantum interference devices (SQUIDs), which measure individual magnetic field components. Consequently, OPMs are employed in a wide range of applications, including biosensing~\cite{bison2003laser,xia2006magnetoencephalography,broser2018optically}, contraband testing~\cite{lee2006subfemtotesla}, and magnetic communications~\cite{gerginov2017prospects,fan2022magnetic,lipka2024multiparameter}. They also aid in dark matter searches~\cite{pospelov2013detecting,afach2021search}, probe fundamental symmetries beyond the energy scales reachable with modern particle colliders~\cite{pendlebury1984search,ayres2021design}, and serve as promising platforms for quantum-enhanced measurements that utilize squeezed light~\cite{wolfgramm2010squeezed,troullinou2023quantum} and macroscopic entanglement~\cite{julsgaard2001experimental,kong2020measurement}.

Accurate vector magnetometry, however, remains a challenge for OPMs due to their inherent scalar operation, as they require an external reference, such as a coil system, to extract directional information. While scalar measurements are often sufficient, many applications benefit from vector measurements of the magnetic field including magnetic anomaly detection~\cite{zhang20224he,prouty2016real}, navigation~\cite{psiaki1991ground,canciani2016absolute}, space exploration~\cite{dougherty2004cassini,korth2016miniature,bennett2021precision}, and geophysics~\cite{leger2009swarm}. For many of these applications accuracy is as crucial as precision. For instance, in planetary magnetosphere studies~\cite{dougherty2004cassini}, enhanced vector accuracy would significantly improve the modeling of internal fields during satellite flybys. Because OPMs require external references to map magnetic field components, vector accuracy is often limited by machining tolerances, drift, and errors in modeling atomic signals.

To date the most common and accurate vector OPMs rely on scalar detection with respect to the magnetic fields produced by a coil system. The vector accuracy of these techniques are generally limited by the calibration accuracy of coil system parameters such as coil factors and coil pair orthogonality. These approaches include a directional varying reference field~\cite{alldredge1960proposed,alexandrov2004three}, fast rotating fields~\cite{wang2023pulsed}, and low-frequency coil modulations~\cite{gravrand2001calibration,andryushkov2022vector}. In particular, coil modulation is a well-established approach that has been implemented with a $^4$He OPM in the European Space Agency SWARM mission~\cite{leger2015flight}. That sensor reached $10$ $\mu$rad directional accuracy after a calibration involving multiple sensor rotations~\cite{gravrand2001calibration}. 

Yet, vector magnetometry via scalar detection combined with a coil system presents certain technical challenges. The first challenge is that the vector sensitivity degrades with increased background magnetic field strength for the same modulation depth. Moreover, in many cases weak modulation fields are necessary due to power requirements, to prevent coupling to external objects, and slew-rate limitations of coil feedback electronics~\cite{zhang2019advanced}. For instance, by employing modulation depths of around 18 $\mu$T, vector component sensitivities down to 0.4 pT$/\sqrt{\text{Hz}}$ have been achieved, as reported in~\cite{wang2023pulsed}. Conversely, in the SWARM mission, vector component sensitivities are restricted to 1 nT$/\sqrt{\text{Hz}}$, a limitation attributed to the smaller modulation depths of 50 nT, as detailed in~\cite{leger2015flight}. Another drawback is that coil modulation techniques offer limited insight from atomic measurements for monitoring environmental drifts in the vector reference (i.e. the coil system). As a result, achieving high accuracy can entail performing iterative calibrations without prior knowledge of whether drift occurred. This process often requires rotations of either the sensor or a large bias field, which can lead to intervals of unnecessary sensor downtime. Finally, a coil system enclosure around the OPM vapor cell increases the physical complexity of the apparatus and limits the overall capabilities for miniaturization~\cite{fairweather1972vector}.

In contrast, using an electromagnetic field as a vector reference, by coupling the atomic ensemble to its polarization structure or propagation direction, offers distinct advantages over coil modulation approaches. In particular, all-optical methods, such as electromagnetically induced transparency (EIT)~\cite{yudin2010vector,gonzalez2024sensitivity}, nonlinear magneto-optical rotation (NMOR)~\cite{pustelny2006nonlinear,meng2023machine}, and methods that detect spin projections on multiple laser beams~\cite{fairweather1972vector,patton2014all,bison2018sensitive,zhang2021vector,petrenko2023all}, are attractive because they enable remote detection, are magnetically quiet, and are compatible with sensor miniaturization. Additionally, radio-frequency fields can be utilized in techniques such as double resonance atomic alignment~\cite{weis2006theory,ingleby2018vector} and the Voigt effect~\cite{pyragius2019voigt}. Unlike coil system modulation techniques, these approaches do not gain vector information from a modulation in the magnetic field strength,  with the exception of~\cite{patton2014all}, but resonantly probe atomic transitions. Thus, by tuning the electromagnetic frequency their vector sensitivity does not degrade at large DC magnetic fields.

Despite these advantages, achieving vector accuracy better than 1-degree (17 mrad) with an electromagnetic reference is a nontrivial task due to challenges in modeling the nonlinear directional dependence of the atomic measurements~\cite{ingleby2018vector,zhang2021vector,meng2023machine,mckelvy2023technical,petrenko2023all}. Furthermore, most reports on vector OPMs using an electromagnetic reference focus solely on sensitivity characterization, leaving vector accuracy underexplored. Major contributing factors to modeling inaccuracy are sensitivity to complex decoherence effects from atomic collisions, imperfect optical pumping, and characterization errors of the electromagnetic reference in terms of spatial inhomogeneity, polarization structure, and time-dependent drifts.

\begin{figure*}[bht]
\centering
\includegraphics[width=1\linewidth]{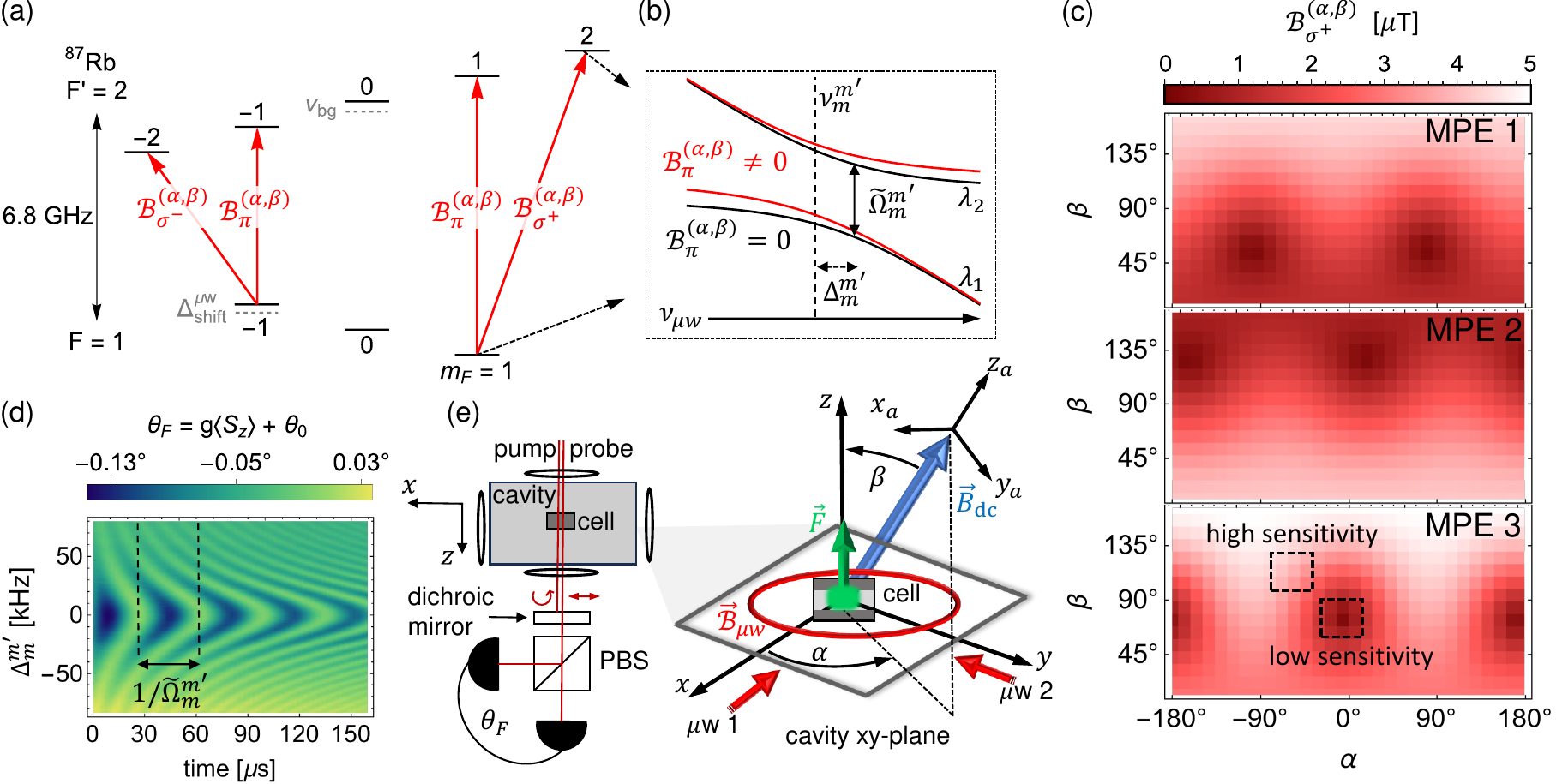}
\caption{Vector magnetometry using Rabi oscillation frequencies referenced to multiple microwave polarization ellipses (MPEs). (a) Energy-level diagram showing the four hyperfine transitions utilized for Rabi detection. Off-resonant microwave driving causes energy-level perturbations ($\Delta^{\mu\text{w}}_{\text{shift}}$) and buffer gas collisions cause a frequency shift in the hyperfine splitting ($\nu_{\text{bg}}$). (b) Generalized Rabi frequencies $\tilde{\Omega}_m^{m^{\prime}}$ are quantified by the difference between dressed-state energies $\lambda_j$ and $\lambda_i$. Off-resonant driving (red) from $\mathcal{B}^{(\alpha,\beta)}_{\pi}$, such as on the $\sigma^+$ transition, causes shifts in these dressed states. (c) The measured $\mathcal{B}_{\sigma^+}^{(\alpha,\beta)}$ spherical microwave component discretely sampled at different magnetic field directions for each of the three MPEs. (d) Measured Rabi-Chevron pattern. (e) The magnetometer apparatus.}
\label{fig:figure1main}
\end{figure*}

In this article, we demonstrate sub-mrad vector magnetometry accuracy through Rabi frequency measurements driven by a microwave field, serving as an electromagnetic vector reference, across multiple atomic transitions. This is achieved through key information in the Rabi measurements to correct for collisional shifts and systematic errors from off-resonant driving, as well as facilitating robust calibration of the microwave field and detection of time-dependent drifts. In this technique, the Rabi measurements derive vector information from the polarization structure of the electromagnetic drive relative to a quantization axis defined by the unknown DC magnetic field. This approach could be extended to a wide range of electromagnetic drives and atomic systems. For example, extension to optical fields could be enabled through two-photon Raman transitions, or to RF fields by coupling magnetic sublevels within a hyperfine manifold.  From a converse perspective, this work also highlights the ability of atomic systems to accurately map microwave field structures, with potential application in microwave circuit characterization~\cite{horsley2015widefield,horsley2016frequency,bohi2012simple}, medical imaging~\cite{nikolova2011microwave,chandra2015opportunities}, and radar~\cite{robinson2021determining}.

Previous experiments using Rabi frequency measurements to map DC electric field components~\cite{koepsell2017measuring} and magnetic field components~\cite{thiele2018self} have only been implemented with ultracold atoms. These experiments showed limited sensitivity and accuracy arising from using small atomic ensembles, employing a single microwave drive, and neglecting effects from off-resonant driving ~\cite{koepsell2017measuring,thiele2018self}. The potential for Rabi measurements to compete with the precision of standard vector magnetometry techniques in vapor-cell platforms is evidenced by the coherence times of Rabi oscillations in vapor cells being similar to those of conventional free-induction decay (FID) measurements~\cite{horsley2013imaging,kiehl2023coherence}. In addition, hyperfine transitions exhibit larger magnetic couplings (e.g., $\mu_{m=1}^{m^{\prime}=2}/h\approx 17$ Hz/nT) than the gyromagnetic ratios $\gamma\approx7$ Hz/nT characterizing Zeeman transitions.

For our implementation, we utilize microwave-driven Rabi oscillations across four hyperfine transitions of $^{87}$Rb [Fig~\ref{fig:figure1main}(a,b)] to calibrate and reference the microwave polarization ellipse (MPE) structure of the driving field to an unknown magnetic field direction $(\alpha,\beta)$. For a given MPE structure ($\vec{\mathcal{B}}^{\mu\text{w}}$), the atomic quantization axis, defined by $(\alpha,\beta)$, gives rise to spherical microwave components $(\mathcal{B}_{\sigma^+}^{(\alpha,\beta)},\mathcal{B}_{\pi}^{(\alpha,\beta)},\mathcal{B}_{\sigma^-}^{(\alpha,\beta)})$ that are linked to the Rabi rates $\Omega_m^{m^{\prime}}$ coupling hyperfine states $\ket{1,m}$ and $\ket{2,m^{\prime}}$ through 
\begin{equation}
\Omega_m^{m^{\prime}}=\mu_m^{m^{\prime}}\mathcal{B}_k^{(\alpha,\beta)}/h,
\label{eq:RabiFormula}
\end{equation}
where $h$ is Planck's constant, $\mu_m^{m^{\prime}}$ is the transition magnetic dipole moment, and $k=\sigma^{\pm},\pi$ denotes hyperfine transitions characterized by $m^{\prime}=m \pm 1$ and $m^{\prime}=m$ respectively. We indirectly detect the Rabi rates, and hence the spherical microwave components, through measurements of the generalized Rabi frequency $(\tilde{\Omega}_m^{m^{\prime}})$ that are well-approximated in the two-level formalism, apart from multi-level effects [Fig.~\ref{fig:figure1main}(a,b)], as
\begin{equation}
\label{eq:twoLevelRabi}
\tilde{\Omega}_m^{m^{\prime}}\approx\sqrt{(\Omega_m^{m^{\prime}})^2+(\Delta_m^{m^{\prime}})^2},
\end{equation}
where $\Delta_m^{m^{\prime}} = \nu_{\mu\text{w}} - \nu_m^{m^{\prime}}$ is the microwave detuning, $\nu_{\mu\text{w}}$ is the microwave frequency, and $\nu_m^{m^{\prime}}$ is the hyperfine transition resonance.

Initially, the MPE structure undergoes calibration using generalized Rabi frequencies $(\tilde{\Omega}_m^{m^{\prime}})$ measured over many known DC magnetic field orientations. Subsequently, these measurements, in conjunction with the directional maps 
[Fig.~\ref{fig:figure1main}(c)] modeled with calibrated MPE parameters, determine the orientation of any magnetic field direction. The use of three distinct MPEs helps mitigate regions where the spherical microwave components, connected to Rabi rates through Eq.~\eqref{eq:RabiFormula}, weakly depend on the magnetic field direction (e.g. $\partial\mathcal{B}_{k}^{(\alpha,\beta)}/\partial \beta \approx 0$) as indicated in Fig.~\ref{fig:figure1main}(c). Utilizing multiple MPE structures that do not all mutually have these regions for the same $(\alpha,\beta)$ enables sensitive vector measurements for all DC field directions.

We account for various systematic shifts in the Rabi oscillation measurements [Fig.~\ref{fig:figure1main}(d)] that affect the accuracy of the directional MPE maps in Fig.~\ref{fig:figure1main}(c). These shifts include the buffer-gas pressure shift ($\nu_{\text{bg}}$), nonlinear Zeeman shifts ($\Delta_{\text{NLZ}}$), and frequency shifts arising from off-resonant driving ($\Delta^{\mu\text{w}}_{\text{shift}}$) [Fig.~\ref{fig:figure1main}(b)]. To correct for these effects, we model them using a full atom-microwave Hamiltonian to fit the Rabi data across multiple atomic transitions. Further, to mitigate systematic errors from microwave drift, we employ running MPE calibrations to track microwave field drifts over a 37.5-minute period, which are consistent with a drift observable derived from the Rabi measurements. We envision that such drift observables, discussed in Sec.~\ref{sec:driftRecal}, could be useful to avoid blind recalibration employed in high-accuracy vector magnetometers~\cite{olsen2003calibration,leger2015flight}. 

Notably, these Rabi measurements use
the same vapor cell parameters, such as temperature and
buffer gas pressure, as those used in sensitive scalar OPMs.
Thus, while the Rabi measurements determine the direction of a DC magnetic field, this approach can be combined with any scalar OPM measurement, such as free-induction decay (FID), to obtain the full magnetic field vector. Even so, we recently showed that Rabi measurements themselves can measure geomagnetic field strengths with improved accuracy over standard OPM measurements from 5 nT down to $0.6$ nT~\cite{kiehl2024correcting}.

We also explore a novel scheme that extracts Rabi rates from resonances during simultaneous Larmor precession and Rabi driving (SPaR). While not as sensitive as Rabi oscillation measurements, SPaR obtains the optimal signal-to-noise ratio (SNR) for DC magnetic field directions where the Rabi oscillation SNR approaches zero. For the Rabi oscillation measurements, this probing deadzone occurs when the magnetic field is perpendicular to the probe beam. The SPaR approach leverages the fact that the Larmor precession signal is maximized at this Rabi probing deadzone. Notably, the existence of a deadzone is a feature to most OPM configurations using a single optical axis~\cite{bloom1962principles}. 

For geomagnetic fields near 50 $\mu$T we achieve 0.54 $\text{nT}/\sqrt{\text{Hz}}$ (11 $\mu$rad/$\sqrt{\text{Hz}}$) vector component (directional) sensitivity and 23 nT (0.46 mrad)  vector component (directional) accuracy with Rabi oscillation measurements away from the Rabi probing deadzone. Using SPaR within the Rabi probing deadzone we achieve 7.3 $\text{nT}/\sqrt{\text{Hz}}$ (150 $\mu$rad/$\sqrt{\text{Hz}}$) sensitivity and 210 nT (4.3 mrad) accuracy. Although there is still room for engineering improvements to optimize microwave stability and reduce measurement deadtime, these results surpass the directional accuracy of many vector OPM techniques using an electromagnetic reference and demonstrate vector component sensitivities comparable to those of high-accuracy OPMs, such as the $^{4}$He OPM used in the SWARM mission~\cite{leger2015flight}.

\section{Apparatus}
\label{sec:Apparatus}
To demonstrate the adaptability of the Rabi technique to compact OPM packages, our setup employs a microfabricated vapor cell ($3\times3\times2$ $\text{mm}^3$) containing $^{87}\text{Rb}$ and $180~\text{torr}$ of $\text{N}_2$ buffer gas. This cell is housed inside a copper rectangular $(4.8\times4.8\times2\text{ cm}^3)$ microwave cavity that, through excitation of two linearly-polarized modes, creates an arbitrarily-shaped MPE in a plane orthogonal to the optical axis at the position of the atoms [Fig.~\ref{fig:figure1main}(e)]. The microwave fields of MPE 1 and MPE 2 are produced by individually driving two microwave ports designed to excite these orthogonally polarized modes. The microwave field for MPE 3 is generated by simultaneously exciting both of these ports, resulting in the microwave field of MPE 3 being effectively approximated by the combined fields of MPE 1 and MPE 2. This cavity is designed with a low quality factor of $Q=62$ (110-MHz linewidth) to minimize frequency dependence of the cavity modes. The cavity temperature is maintained near $100^{\circ}$C using Joule heating from flexible polyimide heaters, which are switched off during measurements. This process, in turn, raises the cell temperature to a similar level. 

Along the cell optical axis, parallel to $\hat{z}$, propagates a 795-nm elliptically polarized pump beam, tuned within a few gigahertz of the $D_1$ line, and a 1-mW probe beam blue-detuned by 170 GHz from the 780-nm $D_2$ line. The pump frequency and polarization are tuned to depopulate the $F = 2$ manifold and enable strong Rabi signals across all hyperfine transitions, while still causing spin polarization for the FID measurement. Complete depopulation of the $F=2$ manifold is limited by the 5.6-GHz optical broadening from Rb-N$_2$ collisions. A polarimeter detects the Faraday rotation $\theta_F=g\langle S_z \rangle+\theta_0$ of the probe beam expressed in terms of the macroscopic z component of the electron spin, a coupling coefficient $g$, and an offset $\theta_0$~\cite{seltzer2008developments}. In this scheme, atomic population dynamics affecting $S_z$ from the Rabi oscillations are nondestructively measured through $\theta_F$. Because the spin dynamics from microwave-driven Rabi oscillations occur along the DC magnetic field ($\vec{B}_{\text{dc}}$), this detection scheme exhibits a deadzone when $\vec{B}_{\text{dc}}$ is oriented perpendicular to the probe beam. The SPaR technique detailed in Sec.~\ref{sec:SPaR} enables Rabi rate measurements even in this probing deadzone. 

The cell and microwave cavity are housed within a three-dimensional (3D) coil system that generates a programmable 50-$\mu$T magnetic field, defined by azimuthal and polar angles $\alpha$ and $\beta$. The coil system is further enclosed within three layers of mu-metal shielding to ensure magnetic isolation from the environment. Calibration of the coil system, detailed in Supplementary 1, defines the laboratory frame $\mathcal{L}=(x,y,z)$ and is essential for the MPE calibrations detailed in Sec.~\ref{sec:MPECalibration}. Notably, outside of the MPE parameter calibration, with calibrated MPE parameters, vector magnetometry using Rabi measurements can be performed in the field independent of DC coil system drifts. 

The DC magnetic field predicted from the coil system calibration also serves as an independent benchmark of the vector accuracy of the Rabi measurements. To account for coil system drift during the 37.5-minute vector data acquisition in our vector assessment in this work, we perform running calibrations of the coil system parameters post-processing from FID measurements described in Supplemental 1. From the FID calibration residuals and the consistency of calibrated coil system parameters across independent calibrations, the directional accuracy of the magnetic field programmed by the coil system, limited by FID heading errors~\cite{kiehl2024correcting} and nonlinearity in the coil current control, is estimated to be within 50 $\mu$rad. This establishes a lower limit on the accuracy with which vector magnetometry from Rabi measurements can be benchmarked. Further details behind the coil system design and calibration are discussed in Supplemental 1. 

\section{Measurement Protocol}
\begin{figure}[!ht]
\centering
\includegraphics[width=.6\linewidth]{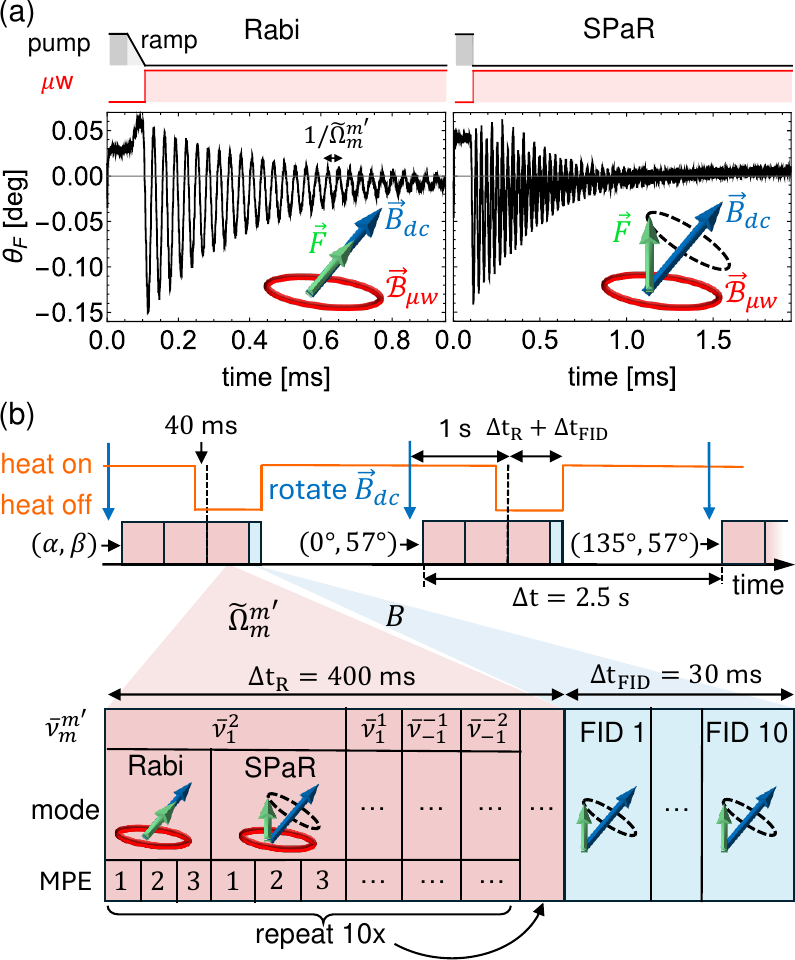}
\caption{(a) Rabi oscillation and SPaR pulse sequences measured at $(\alpha,\beta)=(0\degree,57\degree)$. (b) A timing diagram of the Rabi, SPaR, and FID measurement sequences utilized to benchmark vector accuracy and sensitivity. All Rabi, SPaR, and FID measurements are taken within a $\Delta t_R+\Delta t_{\text{FID}}=430$ ms period. Magnetic field rotations (blue arrow) and deactivation of electrical heating are initiated 1 s and 40 ms before each measurement period to minimize Eddy currents during acquisition. All 12 Rabi and SPaR measurements, corresponding to the three MPE configurations and four hyperfine transitions, are each repeated 10 times.}
\label{fig:figureMeasSeq}
\end{figure}

The measurement protocol for conducting Rabi oscillation and SPaR measurements across the four hyperfine transitions [Fig.~\ref{fig:figure1main}(a)] and three MPE configurations [Fig.~\ref{fig:figure1main}(c)], along with FID measurements, is diagrammed in Fig.~\ref{fig:figureMeasSeq}.
For the Rabi oscillation measurements, we initialize the
atomic ensemble with adiabatic optical
pumping (AOP) to eliminate Larmor precession by aligning the macroscopic atomic spin
along the magnetic field $\vec{B}_{\text{dc}}$ as described in~\cite{kiehl2024correcting}. As depicted in
Fig.~\ref{fig:figureMeasSeq}(a), this involves a 100-mW pump pulse lasting for 50 $\mu$s, followed by a linear ramp-off of the pump power over an additional 50 $\mu$s. In contrast, FID and SPaR measurements utilize a single 100-$\mu$s pulse of pump light at 400 mW to polarize the atomic spins along the pump beam. After pumping, the microwave field from one of the three MPEs is turned on for 0.85 ms and 1.85 ms for Rabi oscillation and SPaR measurements respectively. We drive both of the Rabi and SPaR signals for each hyperfine transition highlighted in Fig.~\ref{fig:figure1main}(a) using a single microwave frequency $\overline{\nu}_m^{m^{\prime}}$ that is near-resonant to the unperturbed hyperfine transition frequency $\nu_{m}^{m^{\prime}}$. The specific microwave frequencies are $\overline{\nu}_{-1}^{-2}=6833.7201$ MHz, $\overline{\nu}_{-1}^{-1}=6834.0701$ MHz, $\overline{\nu}_{1}^{1}=6835.4701$ MHz, and $\overline{\nu}_{1}^{2}=6835.8204$ MHz. 

The Rabi, SPaR, and FID measurements are made over a range of DC magnetic field orientations [Fig.~\ref{fig:figureMeasSeq}(b)]. The sequence of $\vec{B}_{\text{dc}}$ orientations consist of a random direction $(\alpha,\beta)$ interspersed with two repeated field directions used for monitoring microwave drifts. These measurements allow for the evaluation of vector sensitivity and accuracy across all magnetic field directions and enable ongoing MPE calibrations to correct for microwave field drift. At each $\vec{B}_{\text{dc}}$ direction, Rabi oscillation and SPaR measurements for each $\overline{\nu}_m^{m^{\prime}}$ and MPE configuration are repeated 10 times over a $\Delta t_R=400$ ms period followed by 10 repeated FID measurements over a $\Delta t_{\text{FID}}=30$ ms period [Fig.~\ref{fig:figureMeasSeq}(b)]. The 400 ms Rabi measurement sequence is initiated twice before capturing the final Rabi-SPaR-FID sequence to allow time for the microwave components to thermally stabilize.

To prevent Eddy currents during measurements, the cavity heaters are deactivated 40 ms before the final sequence, and the magnetic field rotations are completed 1 second prior. The cooling that occurs due to the deactivation of the cavity heaters during the Rabi-SPaR-FID sequence necessitates a dead time after each sequence to allow the heaters to restore the cavity temperature to $100$ $\degree$C. This dead time restricts the repetition rate of the measurement sequences to $\Delta t=2.5$ seconds and also limits the speed at which MPE calibrations can be performed, as detailed in Sec.~\ref{sec:MPECalibration}. Future systems that limit this deadtime will be even less sensitive to drifts during this calibration period.

\section{Generalized Rabi Frequency Modeling}
\label{sec:RabiModeling}

Generalized Rabi frequencies are influenced by various factors [Fig.~\ref{fig:figure1main}(a,b)], including off-resonant driving, buffer gas collisions, and nonlinear Zeeman shifts, which can constrain the ultimate accuracy of the vector 
measurements. To account for these frequency shifts we model generalized Rabi frequencies in terms of dressed state energies $\lambda_j$ and $\lambda_i$ that correspond to the pair of states coupled by the microwave field [Fig.~\ref{fig:figure1main}(b)]. This relationship is expressed as~\cite{kiehl2024correcting}
\begin{equation}
\label{eq:rabiLambda}
\tilde{\Omega}_m^{m^{\prime}}= \delta \lambda_m^{m^{\prime}} \equiv (\lambda_j-\lambda_i)/h.
\end{equation}
Specifically, $\lambda_j$ and $\lambda_i$ are eigenvalues of the atom-microwave Hamiltonian 
\begin{align}
\begin{split}
\label{eq:Hamiltonian}
H=&\mathcal{M}\big[(A+h\frac{\nu_{\text{bg}}}{2})\mathbf{S}\cdot\mathbf{I}+\mu_B(g_sS_z+g_iI_z)B\big]\mathcal{M}^{\dagger}-\\ &I_2 h\nu_{\mu\text{w}}+\sum_{|m-m^{\prime}|\leq 1}\frac{h}{2}\Big[\ket{\overline{2,m^{\prime}}}\Omega_{m}^{m^{\prime}}\bra{\overline{1,m}}+\text{h.c.}\Big]
\end{split}
\end{align}

The Hamiltonian $H$ in Eq.~\eqref{eq:Hamiltonian} is defined within an atom-frame $\mathcal{A}=(x_a,y_a,z_a)$ where the $z_a$-direction is aligned with the magnetic field direction $(\alpha,\beta)$ with respect to the lab frame $\mathcal{L}$ [Fig.~\ref{fig:figure1main}(e)]. The first line describes the hyperfine and Zeeman interaction and is characterized by electronic $\mathbf{S}=(S_x,S_y,S_z)$ and nuclear $\mathbf{I}=(I_x,I_y,I_z)$ spin operators defined in the $\ket{F,m_F}$ basis, the magnetic dipole hyperfine constant $A$, electron $g_s$ and nuclear $g_i$ Land\'e g-factors, the Bohr magneton $\mu_{B}$, the DC magnetic field strength $B=|\vec{B}_{\text{dc}}|$, and the buffer gas frequency shift $\nu_{\text{bg}}\approx88$ kHz. The magnetic field strength $B\approx50$ $\mu$T is known from FID measurements, and the pressure shift $\nu_{\text{bg}}$ is known from an independent characterization discussed in Supplementary 1. The second line describes the atom-microwave coupling where $I_2$ is the $F=2$ identity operator. We work in a modified hyperfine basis $\ket{\overline{F,m}}=\mathcal{M}\ket{F,m}$ with $\mathcal{M}$ being defined as the operator that diagonalizes the hyperfine and Zeeman terms to preserve NLZ effects during the rotating-wave approximation (RWA).

The spherical microwave components that determine the Rabi rates $\Omega_{m}^{m^{\prime}}$ through Eq.~\ref{eq:RabiFormula} are defined within the atom-frame $\mathcal{A}$ as
\begin{equation}
\label{eq:sphericalMic}
\mathcal{B}_k^{(\alpha,\beta)}=R_y(-\beta)R_z(-\alpha)\vec{\mathcal{B}}\cdot\epsilon_k
\end{equation}
where $\vec{\mathcal{B}}=(\mathcal{B}_xe^{-i\phi_x},\mathcal{B}_ye^{-i\phi_y},\mathcal{B}_z)$ is a complex phasor written in terms of microwave amplitudes $(\mathcal{B}_x,\mathcal{B}_y,\mathcal{B}_z)$ and relative phases $(\phi_x,\phi_y)$ defined in the lab frame $\mathcal{L}$. This phasor describes any microwave field through $\mathcal{B}(t)=\frac{1}{2}[\vec{\mathcal{B}}e^{-i\omega_{\mu\text{w}}t}+\vec{\mathcal{B}}^{\star}e^{i\omega_{\mu\text{w}}t}]$. Spherical projection operators $\epsilon_\pm= \{\frac{1}{\sqrt{2}},\mp \frac{i}{\sqrt{2}},0\}$ and $\epsilon_{\pi}=\{0,0,1\}$ are also defined within the atom-frame $\mathcal{A}$. The phasor $\vec{\mathcal{B}}$ is rotated into the atom-frame through 3D rotation operators $R_{y,z}$ defined with respect to the lab frame $\mathcal{L}$.

For clarity, throughout this article we denote the phasor $\vec{\mathcal{B}}$ and generalized Rabi frequency $\tilde{\Omega}_m^{m^{\prime}}$ corresponding to the $p^{\text{th}}$ MPE configuration, where $p=1,2,3$,  as $(\vec{\mathcal{B}}_m^{m^{\prime}})_p$ and $(\tilde{\Omega}_m^{m^{\prime}})_p$ respectively. This notation also specifies the microwave frequency $\nu_{\mu\text{w}}=\overline{\nu}_m^{m^{\prime}}$ associated with driving a specific hyperfine transition.

\section{MPE Calibration and Long-Term Drift}
\label{sec:MPECalibration}

An essential step to performing vector magnetometry with the Rabi measurements is to calibrate the microwave field parameters.
Because of the microwave frequency dependence of the cavity modes, 12 phasors $(\vec{\mathcal{B}}_m^{m^{\prime}})_p$ must be calibrated, corresponding to 60 microwave parameters. We fit each $(\vec{\mathcal{B}}_m^{m^{\prime}})_p$ using Eq.~\eqref{eq:rabiLambda} over $N=12$ DC field directions by minimizing the following cost function
\begin{equation}
\label{eq:MPEFitting}
(r_m^{m^{\prime}})_p=\sum_{j=1}^{N}(w_m^{m^{\prime}})_p\Big( \delta \lambda_m^{m^{\prime}}((\vec{\mathcal{B}}_m^{m^{\prime}})_p,\alpha,\beta)-(\tilde{\Omega}_m^{m^{\prime}})_p \Big)^2\Big\vert_{(\alpha_j,\beta_j)}.
\end{equation}
We reiterate that during these MPE calibrations and benchmarking, the angles $(\alpha,\beta)$ are determined with an accuracy of $50$ $\mu$rad from running coil system calibrations discussed in Sec.~\ref{sec:Apparatus}. In Eq.~\eqref{eq:MPEFitting} we explicitly denote the magnetic field direction and microwave field dependence of $\delta \lambda_m^{m^{\prime}}$. The generalized Rabi frequencies $(\tilde{\Omega}_m^{m^{\prime}})_p$ and corresponding weights $(w_m^{m^{\prime}})_p=1/(\delta(\tilde{\Omega}_m^{m^{\prime}})_p)^2$ are evaluated from the mean and variance $(\delta (\tilde{\Omega}_m^{m^{\prime}})_p)^2$ of 10 repeated Rabi oscillation measurements.

\begin{figure*}[!t]
\centering
\includegraphics[width=.85\linewidth]{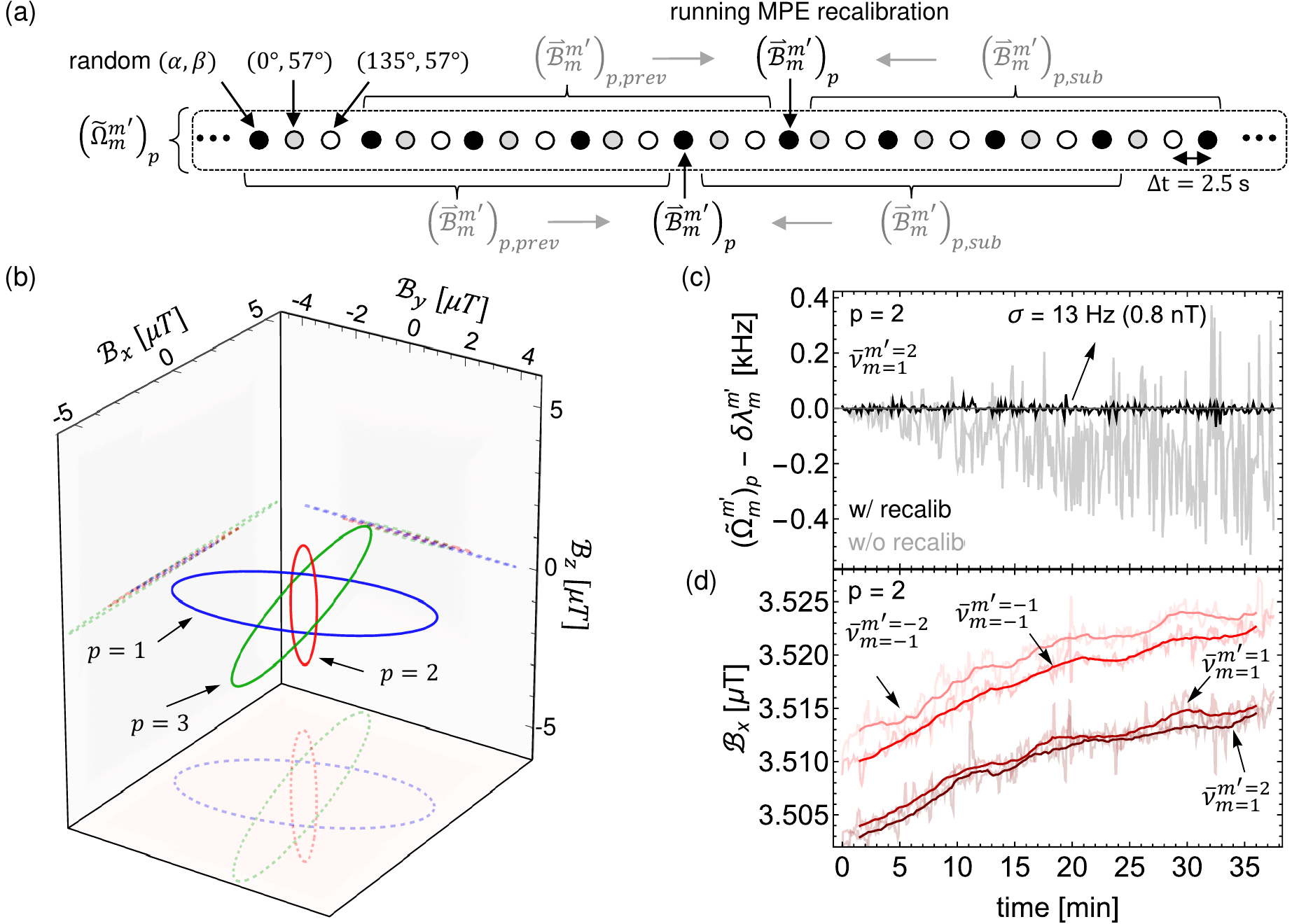}
\caption{(a) The sequence of $\vec{B}_{\text{dc}}$ directions $(\alpha,\beta)$ consisting of random directions (black circles) interspersed with repeated directions $(0\degree,57\degree)$ and $(135\degree,57\degree)$. While all $\vec{B}_{\text{dc}}$ directions in this sequence are utilized in the MPE calibrations, vector measurements are only evaluated at the random $(\alpha,\beta)$. Conversely, the repeated field directions (gray and white circles) are used to consistently monitor microwave drifts. To avoid overfitting, the microwave phasor $(\vec{\mathcal{B}}_m^{m^{\prime}})_p$ at each random $\vec{B}_{\text{dc}}$ direction in this sequence is determined from the weighted average of the two phasors $(\vec{\mathcal{B}}_m^{m^{\prime}})_{p,\text{prev}}$ and $(\vec{\mathcal{B}}_m^{m^{\prime}})_{p,\text{sub}}$ extracted from the MPE calibrations that consist of the $N=12$ directions immediately before and after the random $(\alpha,\beta)$ direction. The calibration length $N$ is limited by the finite time $\Delta t=2.5$ s that is dominated by measurement dead time reserved for cavity heating. (b) Measured MPEs (solid lines) and their projections onto the coordinate planes (dotted lines). The nT-scale variation of the microwave field components arising from the frequency dependence of the cavity modes are not discernible in this plot. (c) Residuals of the $\sigma^+$ generalized Rabi frequency $(\tilde{\Omega}_{m=1}^{m^{\prime}=2})_{p=2}$ at random DC magnetic field directions with and without MPE recalibration. (d) Drift of the $\mathcal{B}_x$ microwave component evaluated at different $\nu_{\mu\text{w}}=\overline{\nu}_m^{m^{\prime}}$. Here, MPE 2 measurements are used as a representative example in (c,d).}
\label{fig:figure3main}
\end{figure*}

To account for MPE drifts, we make Rabi measurements over an $(\alpha,\beta)$ sequence consisting of random $\vec{B}_{\text{dc}}$ directions interspersed with two repeated constant directions [Fig.~\ref{fig:figure3main}(a)]. At each random $(\alpha,\beta)$, we estimate $(\vec{\mathcal{B}}_m^{m^{\prime}})_p$ by using Rabi measurements taken at the 12 previous directions to calibrate a phasor $(\vec{\mathcal{B}}_m^{m^{\prime}})_{p,\text{prev}}$ with squared residual error (Eq.~\eqref{eq:MPEFitting}) $(r_m^{m^{\prime}})_{p,\text{prev}}$ and the subsequent 12 directions to calibrate $(\vec{\mathcal{B}}_m^{m^{\prime}})_{p,\text{sub}}$ with the corresponding error $(r_m^{m^{\prime}})_{p,\text{sub}}$, and calculate the weighted average 
\begin{equation}(\vec{\mathcal{B}}_m^{m^{\prime}})_p=\frac{((r_m^{m^{\prime}})_{p,\text{sub}}(\vec{\mathcal{B}}_m^{m^{\prime}})_{p,\text{prev}}+(r_m^{m^{\prime}})_{p,\text{prev}}(\vec{\mathcal{B}}_m^{m^{\prime}})_{p,\text{sub}})}{((r_m^{m^{\prime}})_{p,\text{prev}}+(r_m^{m^{\prime}})_{p,\text{sub}})}.
\end{equation}
This approach avoids overfitting by ensuring that the calibrated MPE parameters in $(\vec{\mathcal{B}}_m^{m^{\prime}})_{p}$ are validated exclusively against Rabi measurements taken from magnetic field directions not used during the calibration process.

Over 37.5 minutes, we observe nT-scale drifts of the microwave field components and mrad-scale drifts in the phases as exemplified for MPE 2 in Fig.~\ref{fig:figure3main}(d). The microwave field drift is dominated by the contraction of the microwave cavity modes due to a few $\degree$C cooling of the microwave cavity. Details on monitoring the temperature drift from the Rabi dephasing rate and accounting for temperature-dependent shifts in $\nu_{\text{bg}}$ are discussed in Supplementary 1. For these calibrations, we found that it was sufficient to recalibrate only the relative phase $\delta\phi=\phi_y-\phi_x$ by leaving $\phi_x$ fixed after an initial calibration instead of resolving for both $\phi_x$ and $\phi_y$. We note that most of this 37.5 minute duration is spent heating the cavity and not taking measurements. Calibrating all MPE parameters across 12 random $\vec{B}_{\text{dc}}$ directions currently requires 132 seconds. A promising approach to eliminate this duration in future work involves the use of continuous laser heating~\cite{mhaskar2012low}, which would not interfere with atomic measurements. By disregarding technical deadtime, the calibration time could be reduced to 1.8 seconds, as the total measurement duration for Rabi oscillation data is 150 ms. This duration could be further shortened to a few hundred milliseconds by reducing the MPE model parameters, for instance, by eliminating the frequency dependence of the cavity modes with improved microwave engineering. Shortening the calibration time would also help minimize systematic errors caused by potential drifts in the DC background magnetic field during MPE calibration.

To assess the calibration accuracy, we show residuals of $\sigma^+$ generalized Rabi frequencies $(\tilde{\Omega}_{m=1}^{m^{\prime}=2})_{p=2}$ against the $\delta \lambda_{m=1}^{m^{\prime}=2}$ model utilizing recalibrated MPE 2 parameters in Fig.~\ref{fig:figure3main}(c) as a representative example. The residuals are only evaluated over random magnetic field orientations. Over 37.5 minutes these residuals exhibit a standard deviation of $\sigma=13$ Hz, which corresponds to $\sigma=0.8\text{ nT}$  fluctuations for the spherical microwave component $\mathcal{B}_{\sigma^+}^{(\alpha,\beta)}$. These fluctuations are approximately three times greater than the anticipated $\sigma=5$ Hz ($0.3$ nT), a figure expected due to the limited calibration length of $N=12$ (see Supplement 1). This discrepancy suggests that the systematic errors in the calibrated MPE parameters are estimated to be $\delta \mathcal{B}_x=0.4$ nT, $\delta\mathcal{B}_y=0.3$ nT, $\delta\mathcal{B}_z=0.2$ nT, and $\delta (\phi_y-\phi_x)=0.3$ mrad. These calibration errors are consistent with the directional accuracies deduced in the vector magnetometry evaluation in Sec.~\ref{sec:vecMag}.

\section{Simultaneous Precession and Rabi (SP\MakeLowercase{a}R)}
\label{sec:SPaR}

\begin{figure*}[!hbt]
\centering
\includegraphics[width=1\linewidth]{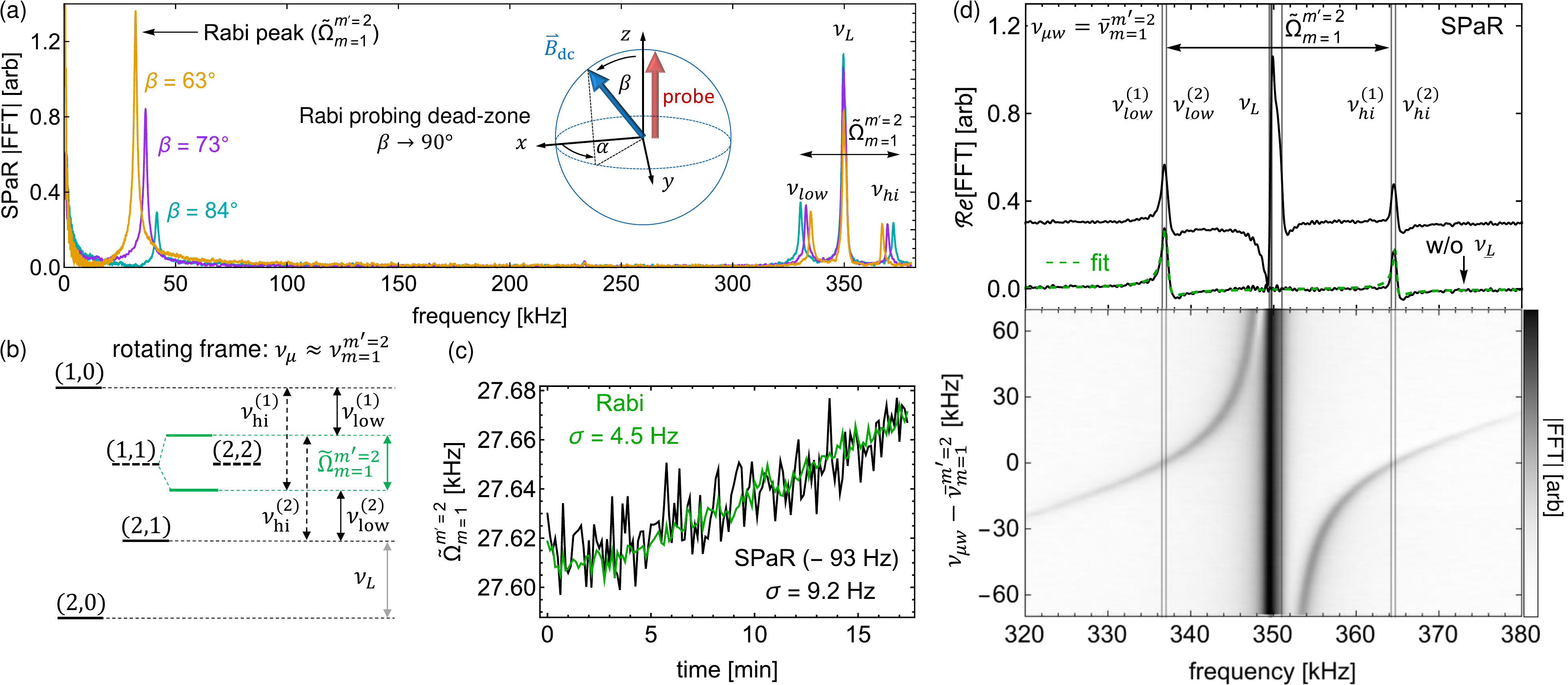}
\caption{Rabi rate detection using SPaR. (a) Directional dependence of Rabi oscillation and SPaR signal amplitude. This plot shows the fast Fourier transform (FFT) of SPaR time-domain signals like the one shown in Fig.~\ref{fig:figureMeasSeq}(a). (b) Energy-level diagram in the rotating frame near the $\sigma^+$ transition frequency. Coupling between microwave dressed states (green) with Zeeman sublevels (black) are indicated by double-sided arrows that indicate $\nu_{\text{hi}}$ (long, dashed) and $\nu_{\text{low}}$ (short, solid) frequency components. (c) Generalized Rabi frequency measurements using Rabi oscillations (green) and SPaR (black). SPaR measurements are shifted by the values in parenthesis. (d) Measured SPaR spectrum for the $\sigma^+$ transition. The top plot shows the real part of the SPaR FFT at the $\sigma^+$ transition frequency $\nu_{\mu\text{w}}=\overline{\nu}_{m=1}^{m^{\prime}=2}$, both with and without the broadened Larmor resonances occuring at $\nu_{\text{L}}$. A green dashed line represents a fit applied to the SPaR FFT after removing the Larmor resonances. The bottom plot shows the absolute value of the SPaR FFT over a range of microwave frequencies centered about $\nu_{\mu \text{w}}=\overline{\nu}_{m=1}^{m^{\prime}=2}$, with the broadened Larmor resonances retained. Vertical lines indicate $\nu_{\text{low}}$, $\nu_{\text{hi}}$, and $\nu_{\text{L}}$ precessional frequencies calculated from the eigenstates of $H$.}
\label{fig:figure2main}
\end{figure*}

For the single optical-axis geometry utilized in this work, Rabi oscillations, and corresponding vector measurements, exhibit a deadzone when $\beta \rightarrow 90\degree$ [Fig.~\ref{fig:figure2main}(a)]. In this section we discuss how SPaR is an effective method to detect Rabi rates $\Omega_m^{m^{\prime}}$ within this probing deadzone through precessional frequency components that appear around the Larmor frequency $\nu_L$ labeled $\nu_{\text{low}}$ and $\nu_{\text{hi}}$ in Fig.~\ref{fig:figure2main}(a). Like FID measurements, the amplitude of these SPaR peaks are maximal when $\beta=90\degree$. 

These SPaR peaks occur due to couplings between the microwave-dressed states and the surrounding Zeeman sublevels, which are conceptually diagrammed in Fig.~\ref{fig:figure2main}(b) from the perspective of a rotating reference frame near the $\sigma^+$ transition frequency $\nu_{\mu\text{w}}=\overline{\nu}_{m=1}^{m^{\prime}=2}$. Akin to the Rabi modeling described in Sec.~\ref{sec:RabiModeling}, the SPaR frequency components are modeled from the differences between pairs of eigenvalues of $H$, depicted by arrows diagrammed in Fig.~\ref{fig:figure2main}(b). This diagram illustrates that the frequency separations of the $\nu_\text{hi}$ and $\nu_\text{low}$ components closely approximate the $\sigma^+$ Rabi frequency. The eigenvalue analysis with known magnetic and microwave-field parameters indicates that each of the SPaR peaks, labeled by $j\in(\text{low, hi})$, consist of two frequency components $\nu_{\text{j}}^{(1)}$ and $\nu_{\text{j}}^{(2)}$ displayed as vertical lines in the $\sigma^+$ SPaR spectrum shown in Fig.~\ref{fig:figure2main}(d). The $\sigma^-$ transition, also consisting of the stretched $|m^{\prime}|=2$ sublevel, has a SPaR spectrum that is qualitatively the same. SPaR measurements and modeling details for the $\pi$ transitions are further discussed in Supplement 1. 

To fit Rabi rates $\Omega_m^{m^{\prime}}$ with SPaR, we begin by estimating the three spherical microwave components $\mathcal{B}_k^{(\alpha,\beta)}$, defined in Eq.~\eqref{eq:sphericalMic}, from the $\sigma^{+}$ and $\sigma^{-}$ SPaR spectra.  We do this by first estimating the centers of the $\nu_{\text{low}}$ and $\nu_{\text{hi}}$ peaks in the SPaR FFT signal using a peak-finding algorithm. Then, we approximate $\nu_{\text{hi}}-\nu_{\text{low}}\approx \Omega_m^{m^{\prime}}$ along with Eq.~\eqref{eq:RabiFormula} to estimate $\mathcal{B}_{\sigma^{\pm}}^{(\alpha,\beta)}$. From the total microwave field amplitude $|\vec{\mathcal{B}}|$, known from MPE calibrations described in Sec.~\ref{sec:MPECalibration}, $\mathcal{B}_{\pi}^{(\alpha,\beta)}$ is calculated using $|\mathcal{B}_{\pi}^{(\alpha,\beta)}|^2=|\vec{\mathcal{B}}|^2-|\mathcal{B}_{\sigma^{+}}^{(\alpha,\beta)}|^2-|\mathcal{B}_{\sigma^{-}}^{(\alpha,\beta)}|^2$. By assuming that the spherical components are positive instead of complex numbers, we can estimate all Rabi rates in Eq.~\eqref{eq:Hamiltonian} with 1 kHz accuracy, and further, estimate all eigenvalues $\lambda_j$ of $H$. We linearize these eigenvalues about these estimates to enable faster computation during SPaR fitting that is described in Supplement 1.

Next, we remove the broadened Larmor resonance $\nu_L \approx 350$ kHz appearing in each SPaR spectra (Supplement 1) and use the following fitting equation to fit the $\sigma^{\pm}$ SPaR spectra
\begin{equation}
\label{eq:FSP}
\mathcal{R}e[\text{FFT}] = \sum_{j\in \{\text{low,hi} \} }a_j\frac{\text{cos}[\phi_j]-\text{sin}[\phi_j](f-\overline{\nu}_j+f_{\text{shift}})}{(f-\overline{\nu}_{j}+f_{\text{shift}})^2+w_j^2/4}
\end{equation}
where $\overline{\nu}_{j}$ is the mean frequency of the two $\nu_{j}^{(1)}$ and $\nu_{j}^{(2)}$ frequency components, $f_{\text{shift}}$ is a phenomenological frequency shift, $\phi_{j}$ are phase shifts, and the strength and broadening of this signal is given by amplitudes $a_j$ and linewidths $w_j\approx 2$ kHz. We use the mean frequency $\overline{\nu}_{j}$  instead of precisely modeling the lineshape of each of the $\nu_{j}^{(1)}$ and $\nu_{j}^{(2)}$ to avoid overfitting. During the $\sigma^+$ SPaR fitting, $\mathcal{B}_{\sigma^+}^{(\alpha,\beta)}$ is the only free parameter, where $\mathcal{B}_{\sigma^-}^{(\alpha,\beta)}$ and $\mathcal{B}_{\pi}^{(\alpha,\beta)}$ are held at their initial estimates. Correspondingly, $\mathcal{B}_{\sigma^-}^{(\alpha,\beta)}$ and $\mathcal{B}_{\pi}^{(\alpha,\beta)}$ are the only free parameters during SPaR fitting of the $\sigma^-$ and $\pi$ transitions respectively.

The accuracy of the SPaR Rabi fits is within a few hundred Hz [Fig.~\ref{fig:figure2main}(c)] due to uncertainty in the relative amplitudes of the unresolved frequency components that make up $\nu_{\text{low}}$ and $\nu_{\text{hi}}$. Despite this, over a 17-minute period, we observe changes in Rabi rates using SPaR that align with those measured from Rabi oscillations as illustrated in Fig.~\ref{fig:figure2main}(c). These observations occur at a magnetic field orientation $(\alpha,\beta)=(135\degree,57\degree)$, which is intermediary to the Rabi and SPaR deadzones.

\section{Vector Magnetometry Evaluation}
\label{sec:vecMag}

\begin{figure*}[!bth]
\centering
\includegraphics[width=.95\linewidth]{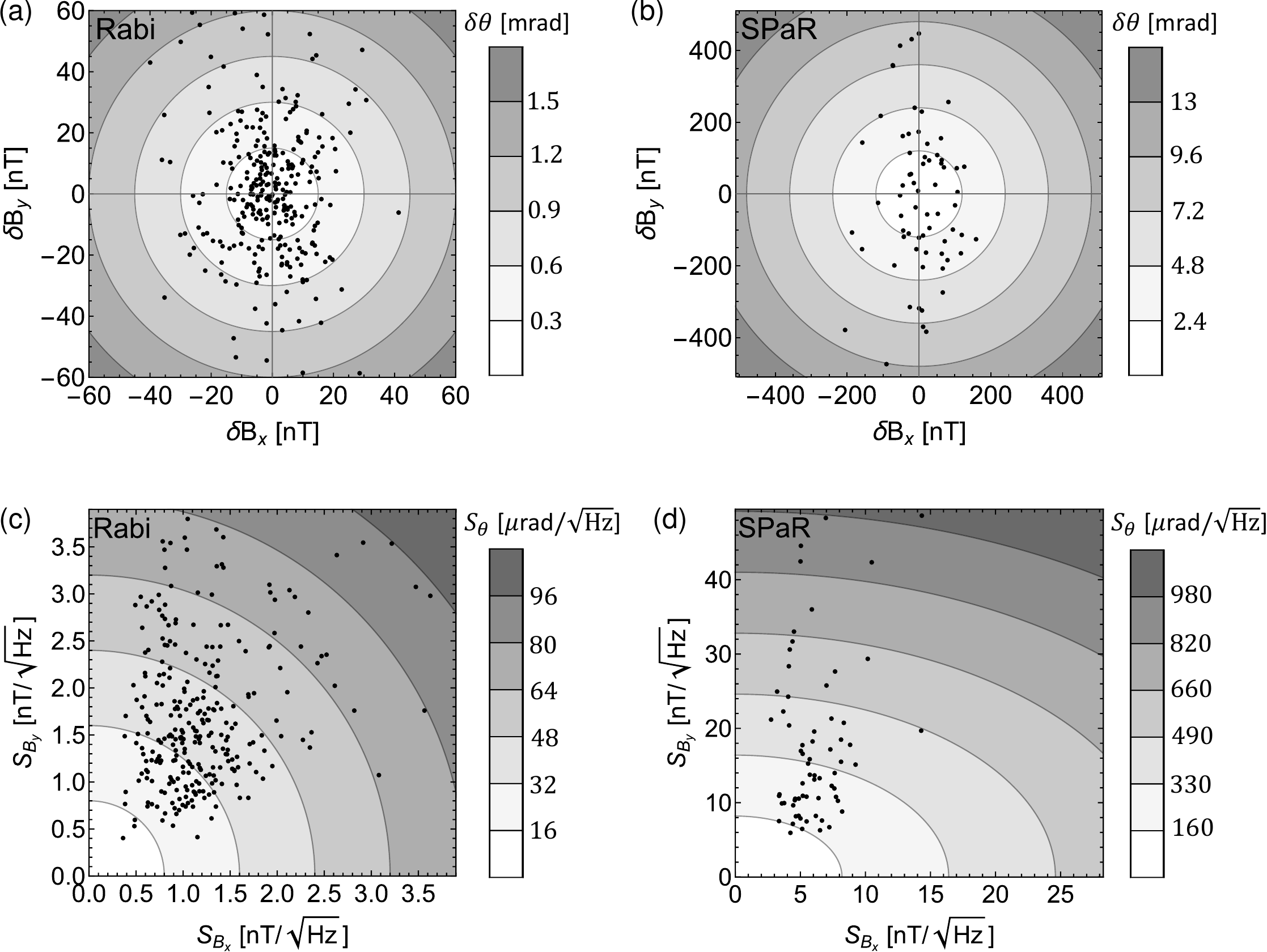}
\caption{Analysis of vector accuracy and sensitivity using either Rabi oscillation and SPaR measurements. (a) and (b) show the vector accuracy, and (c) and (d) show the vector sensitivity for Rabi and SPaR techniques evaluated over 300 and 80 random field directions respectively.}
\label{fig:figure5main}
\end{figure*}

The evaluation of the accuracy and sensitivity of the vector measurements, conducted over 300 random $\vec{B}_{\text{dc}}$ orientations, are depicted in Fig.~\ref{fig:figure5main}. In the case of Rabi oscillation data, we exclude field directions that have polar angles within $90\degree\pm 5 \degree$, whereas for SPaR, the evaluation is specifically focused on polar angles that fall within $90\degree\pm 10 \degree$. Despite the periodicity of $\mathcal{B}^{(\alpha,\beta)}_k$ with respect to DC magnetic field direction $(\alpha,\beta)$ [Fig.~\ref{fig:figure1main}(c)], the Rabi measurements across three MPEs find unique solutions for $(\alpha,\beta)$. See Supplemental 1 for further discussion on the uniqueness of the field orientation deduced from the Rabi measurements. 

For the Rabi oscillation data, we solve for the DC magnetic field direction $(\alpha,\beta)$ by fitting the generalized Rabi frequency $(\tilde{\Omega}_m^{m^{\prime}})_s$ to the eigenvalue model $\delta \lambda_m^{m^{\prime}}$ with the cost function
\begin{equation}
\label{eq:rabiMinimize}
r_{\text{rabi}}^{(\alpha,\beta)}=
\sum_{p=1}^3 \sum_{m,m^{\prime}}(\tilde{w}_m^{m^{\prime}})_p \Big( \delta \lambda_m^{m^{\prime}}((\vec{\mathcal{B}}_m^{m^{\prime}})_p,\alpha,\beta)-(\tilde{\Omega}_m^{m^{\prime}})_p \Big)^2.
\end{equation}
We note that the microwave parameters $(\vec{\mathcal{B}}_m^{m^{\prime}})_p$ are obtained from the calibration described in Sec.~\ref{sec:MPECalibration}. For the SPaR data, we solve for $(\alpha,\beta)$ by fitting the measured Rabi rate $(\Omega_m^{m^{\prime}})_p$ to the spherical microwave component (Eq.~\eqref{eq:sphericalMic}) with the cost function
\begin{align}
 \label{eq:sparMinimize}
\begin{split}
r_{\text{spar}}^{(\alpha,\beta)}&=\sum_{p=1}^3 \sum_{m,m^{\prime}}(w_m^{m^{\prime}})_p \times\\&
 \Big(\Big|R_y(-\beta)R_z(-\alpha)\frac{\mu_m^{m^{\prime}} (\vec{\mathcal{B}}_m^{m^{\prime}})_p}{h}\cdot\epsilon_k\Big| -(\Omega_m^{m^{\prime}})_p\Big)^2.
\end{split}
\end{align}
The absolute value in Eq.~\eqref{eq:sparMinimize} emphasizes the approximation, only during SPaR fitting, that the spherical microwave components and Rabi rates are positive numbers. For both, Eq.~\eqref{eq:rabiMinimize} and Eq.~\eqref{eq:sparMinimize}, the pairs of magnetic sublevels $m,m^{\prime}$ are only summed over the four hyperfine transitions in Fig.~\ref{fig:figure1main}(a). The weights $(\tilde{w}_m^{m^{\prime}})_p=(\delta (\tilde{\Omega}_m^{m^{\prime}})_p)^{-2}((r_m^{m^{\prime}})_p)^{-1}$ in Eq.~\eqref{eq:rabiMinimize} are given by the fitting errors $\delta (\tilde{\Omega}_m^{m^{\prime}})_p$ of the generalized Rabi frequencies and further weighted by the MPE squared residual error $\overline{r}_m^{m^{\prime}}=((r_m^{m^{\prime}})_{p,1}+(r_m^{m^{\prime}})_{p,2})/2$, defined in terms of $(r_m^{m^{\prime}})_{p,1}$ and $(r_m^{m^{\prime}})_{p,2}$ introduced in Sec.~\ref{sec:MPECalibration}. The weights $(w_m^{m^{\prime}})_p$ in Eq.~\eqref{eq:sparMinimize} are similarly defined.

To avoid issues arising from $\alpha$ becoming undefined as $\beta\rightarrow 0$, accuracy and sensitivity evaluations are conducted by mathematically rotating the DC magnetic field $\vec{B}_m$ onto the z-axis of the lab frame $\mathcal{L}$ through
\begin{equation}
\label{eq:MagRotate}
(\delta B_x,\delta B_y, B_z)=R_y(-\beta)R_z(-\alpha)\vec{B}_m.
\end{equation}
The rotation in Eq.~\eqref{eq:MagRotate} utilizes the field direction $(\alpha,\beta)$ predicted by a coil system calibration with 50 $\mu$rad accuracy (see Supplement 1). As mentioned in Sec.~\ref{sec:Apparatus}, running coil system calibrations are used throughout the vector measurements to eliminate errors due to coil system drifts. Up to the accuracy of the coil system calibration, residual transverse components $\delta B_x$ and $\delta B_y$ are due to vector inaccuracies. These transverse components are converted into a directional accuracy through 
\begin{equation}
\label{eq:angularEval}
\delta \theta=\text{tan}^{-1}\Big(\frac{1}{B}\sqrt{\delta B_x^2+\delta B_y^2}\Big),
\end{equation}
where $B=|\vec{B}_m|$.
This directional accuracy is represented by the shaded circular arcs in Fig.~\ref{fig:figure5main}(a,b).

The mean directional accuracies of the Rabi oscillation and SPaR vector data are $0.46$ mrad ($0.026 ^{\circ}$) and $4.3$ mrad ($0.25 ^{\circ}$) respectively. These accuracies are consistent with the measured MPE calibration residuals from Rabi oscillation measurements at the 10 Hz scale [Fig.~\ref{fig:figure3main}(c)] and the 100 Hz scale systematics in the SPaR measurements [Fig.~\ref{fig:figure2main}(d)]. Rabi frequency systematic errors at the 10 Hz scale may partially be due to imperfect microwave drift compensation, but can also be due to a variety of sources stemming from microwave phase noise, DC magnetic field spatial gradients, and microwave spatial inhomogeneity discussed further in Supplement 1.

We characterize the vector component sensitivity in terms of the fluctuations of $\delta B_{x}$ and $\delta B_{x}$ in Eq.~\eqref{eq:MagRotate} through
\begin{equation} S_{B_i} = \sigma_{B_i}\sqrt{2t_m},
\end{equation}
where $i = x, y$, $\sigma_{B_i}$ is the standard error evaluated from 10 repeated measurements of the $\delta B_i$ components, and $t_m$ is the total measurement time. For the Rabi oscillation data, the total measurement duration is $t_m=150$ ms, while for SPaR measurements the duration is $t_m=270$ ms. Included in each $t_m$ is the FID measurement time lasting 30 ms. The directional sensitivity is evaluated in terms of the geometric mean of $S_{B_x}$ and $S_{B_y}$ through
\begin{equation}
S_{\theta}=\text{tan}^{-1}\Big(\frac{1}{B}\sqrt{\delta S_{B_x}^2+S_{B_y}^2}\Big).
\end{equation}
With this sensitivity metric, the Rabi oscillation vector measurements reach a mean directional sensitivity of $49$ $\mu$rad/$\sqrt{\text{Hz}}$ evaluated over random field directions, with the best sensitivity reaching $11$ $\mu$rad/$\sqrt{\text{Hz}}$. As examined in Supplemental 1, the measured directional sensitivity is about three times worse than expected if the noise in the Rabi measurements were purely due to polarimeter noise. Suboptimal performance below this detector limit is attributed to shot-to-shot amplitude fluctuations of the microwave field. A summary of the accuracy and sensitivity metrics for the Rabi oscillation and SPaR data are reported in Table~\ref{tab:vectorMetrics}.

As a sensitivity and bandwidth comparison, the vector component sensitivities reported in Table~\ref{tab:vectorMetrics} are at the same 1 nT$/\sqrt{\text{Hz}}$ level reached in the Helium SWARM mission ~\cite{gravrand2001calibration}.  With a measurement duration of \(t_m = 150\) ms, the vector magnetometer would achieve a bandwidth of 3.3 Hz if not for the currently implemented deadtime required for cavity heating and eddy current dissipation. For perspective, the vector mode of the ESA Swarm $^{4}$He magnetometer, operating in near-Earth orbit, has a bandwidth of 0.4 Hz~\cite{leger2009swarm}. While this is not a direct comparison due to a variety of apparatus differences such as the smaller cell volume employed in this work, it demonstrates comparable vector sensitivity and bandwidth to high accuracy OPM implementations. Future advancements could include implementing full 3D control of the microwave field, moving beyond planar MPEs. This would enable the strategic selection of MPEs to optimize vector sensitivity across all field directions. Additionally, as outlined in Sec.~\ref{sec:MPECalibration}, continuous laser heating~\cite{mhaskar2012low} offers a promising approach to eliminating deadtime associated with cavity heating. Replacing the microwave cavity with a dielectric resonator platform is another viable path to mitigate eddy current effects while reducing the sensor's overall size and weight.

\begin{table}[htbp]
\centering
\caption{A summary of the accuracy and sensitivity metrics for the Rabi oscillation and SPaR measurements. The first two rows detail the vector component and directional accuracy, while the last four rows describe vector component and directional sensitivity. Brackets $\langle ... \rangle$ indicate an average over random DC magnetic field directions, while min$(...)$ represents the minimum measured value.}
\begin{tabular}{ccc}
\hline
Vector Metric & Rabi & SPaR  \\
\hline
$\Big \langle \sqrt{\delta B_x^2+\delta B_y^2} \Big \rangle$ & 23 nT & 210 nT \\
$\big \langle \delta \theta \big \rangle$ & 0.46 mrad & 4.3 mrad \\
$\Big\langle \sqrt{\delta S_{B_x}^2+\delta S_{B_y}^2} \Big\rangle$ & 2.5 nT/$\sqrt{\text{Hz}}$ & 25 nT/$\sqrt{\text{Hz}}$\\
$\big \langle S_{\theta} \big \rangle$ & 49 $\mu$rad/$\sqrt{\text{Hz}}$ & 490 nT/$\sqrt{\text{Hz}}$\\
min$\Big(\sqrt{\delta S_{B_x}^2+\delta S_{B_y}^2} \Big)$ & 0.54 nT/$\sqrt{\text{Hz}}$ & 7.3 nT/$\sqrt{\text{Hz}}$\\
min$\big(S_{\theta} \big)$ & 11 $\mu$rad/$\sqrt{\text{Hz}}$ & 150 $\mu$rad/$\sqrt{\text{Hz}}$\\
\hline
\end{tabular}
  \label{tab:vectorMetrics}
\end{table}

\section{MPE Drift Detection Without Recalibration}
\label{sec:driftRecal}

Running MPE calibrations described in Sec.~\ref{sec:MPECalibration} are used to compensate any microwave field drifts for the purposes of benchmarking vector magnetometry errors. In fact, all types of vector OPMs will inevitably experience drift in their reference system, whether defined by a coil system or an electromagnetic field, over extended periods. Current vector OPM techniques, however, have limited information from atomic measurements to detect sensor drift. For example, vector OPMs using coil modulations can only compare consistency between the norm of the measured vector components $(B_x,B_y,B_z)$ with the scalar value $B$ to detect drift in parameters like coil non-orthogonality and coil factors. As a result, recalibration of the vector reference is often performed blindly, regardless of whether the drift occurred, causing additional downtime in the sensor acquisition.

In contrast to vector OPMs based on scalar detection, Rabi measurements across multiple transitions and MPEs contain much more independent information to detect drift in the MPE reference.  For example, in this work there are nine independent Rabi measurements corresponding to the three spherical microwave components of MPE 1, MPE 2, and MPE 3. In comparison, if the frequency dependence of the microwave cavity modes is neglected, the sensor is characterized by 15 MPE parameters and two angles defining the DC magnetic field direction. However, because the microwave field corresponding to MPE 3 is the sum of the fields of MPE 1 and MPE 2, the number of independent MPE parameters can be further reduced. In this case, the number of parameters decreases from 15 to 11. This situation is expressed as 
\begin{align}
\label{eq:sumMPEs}
\begin{split}
\vec{\mathcal{B}}_3=&\vec{\mathcal{B}}_1+\vec{\mathcal{B}}_2=\{\mathcal{B}_{x,1}e^{-i\phi_{x,1}},\mathcal{B}_{y,1}e^{-i\phi_{y,1}},\mathcal{B}_{z,1}\}+\\
&\{\mathcal{B}_{x,2}e^{-i(\phi_{x,2}+\phi_r)},\mathcal{B}_{y,2}e^{-i(\phi_{y,2}+\phi_r)},\mathcal{B}_{z,2}e^{-i\phi_{r}}\},
\end{split}
\end{align}
where the additional parameter $\phi_r$ accounts for the relative phase between the microwave fields of MPE 1 and MPE 2. This is further demonstrated in Supplementary 1, where the independently calibrated MPE 3 parameters at a fixed microwave frequency are well approximated by the 10 microwave parameters of MPE 1 and MPE 2, along with a relative phase of $\phi_r=1.058$ rad, using Eq.~\eqref{eq:sumMPEs}. Thus, without frequency dependence of the cavity modes, the nine independent Rabi measurements are close to the 13 parameters (11 microwave and two angles defining the DC field orientation) that characterize the vector magnetometer. Two of these parameters will always be undefined without additional external information because any vector magnetometer cannot distinguish between global sensor rotations and DC magnetic field rotations.

\begin{figure}[!ht]
\centering
\includegraphics[width=.6\linewidth]{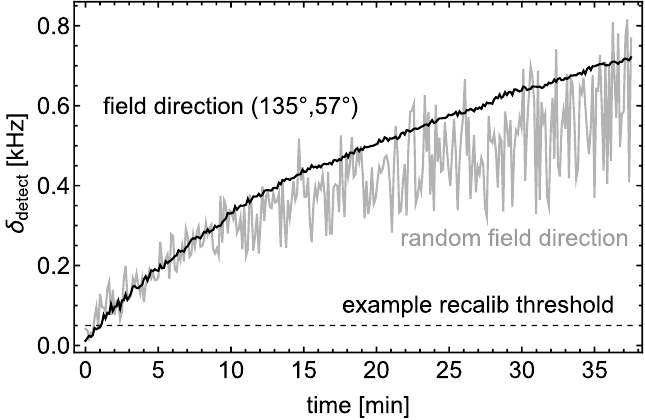}
\caption{The drift parameter $\delta_{\text{detect}}$ evaluated over 37.5 minutes demonstrating drift detection without full MPE recalibration.}
\label{fig:figure4main}
\end{figure}

As a proof of concept, Fig.~\ref{fig:figure4main} shows how the RMS error $\delta_{\text{detect}}=(r_{\text{rabi}}^{(\alpha,\beta)})^{1/2}$, computed from Eq.~\eqref{eq:rabiMinimize} after solving for $(\alpha,\beta)$ and setting the weights $(\tilde{w}_m^{m^{\prime}})_s \rightarrow 1$, provides a drift-detection measure. Establishing a specific threshold for $\delta_{\text{detect}}$ allows it to act as a cue to recalibrate, and thus eliminating the necessity for blind recalibration. Exploring how to further utilize MPE drift measures like $\delta_{\text{drift}}$ to mitigate MPE recalibration is the subject of future work.

\section{Conclusion}

This work demonstrates a vector OPM utilizing a microwave vector reference that achieves sub-milliradian accuracy [see Table~\ref{tab:vectorMetrics}] and a vector component sensitivity as low as 0.54 nT$/\sqrt{\text{Hz}}$ (11 $\mu$rad$/\sqrt{\text{Hz}}$)  at magnetic field strengths near $50$ $\mu$T. This accuracy is reached by leveraging information in Rabi measurements across multiple hyperfine transitions to correct for systematic shifts arising from buffer gas collisions, nonlinear Zeeman effects, and off-resonant microwave driving. In addition, by using simultaneous precession and Rabi driving (SPaR), we also demonstrated deadzone-free vector measurements, even with the use of a single optical axis. Furthermore, utilizing three unique MPE configurations eliminated insensitive regions in the directional maps of the spherical microwave components [Fig.~\ref{fig:figure1main}(c)] and enabled a unique measurement of the magnetic field direction, despite the periodicity of the spherical microwave components $\mathcal{B}_k^{(\alpha,\beta)}$ with respect to $(\alpha,\beta)$.

In future work, using magnetically quiet techniques like laser heating~\cite{mhaskar2012low} to heat the vapor cell, rather than the entire microwave cavity, could eliminate measurement downtime and improve temperature stability, leading to more accurate MPE calibrations and longer durations of vector measurements before MPE drift errors become significant. Furthermore, designing a microwave system that does not depend on the microwave frequency, combined with full 3D control of the microwave field, could allow drift-tracking parameters derived from Rabi measurements (as discussed in Sec.~\ref{sec:driftRecal}) to detect and correct for microwave drifts, updating MPE parameters without the need for full recalibration via field rotations from the DC coil system.

The methodologies introduced here also lay the groundwork towards extending the Rabi vector measurements beyond its current scope. For instance, the techniques presented in Sec.~\ref{sec:MPECalibration} for calibrating the polarization ellipse of a microwave field could be useful for characterizing the accuracy of other microwave sensors based on atomic vapors such as atomic candle~\cite{coffer2000atomic,kinoshita2010atomic,tretiakov2019microwave} and Rydberg EIT~\cite{mohapatra2007coherent,sedlacek2012microwave,holloway2014broadband} methods. The Rabi vector technique could also be adapted with different electromagnetic references, such as radio-frequency fields that couple Zeeman transitions within a hyperfine manifold, commonly used in M$_z$ and M$_x$ configurations~\cite{alexandrov2003recent}. Alternatively, an all-optical approach could be explored by driving Rabi oscillations through two-photon Raman transitions. Furthermore, extending the model in Sec.~\ref{sec:RabiModeling} to also incorporate the amplitude of Rabi oscillations could be applied to the probing dead zone behavior [see Fig.~\ref{fig:figure2main}] to accurately align the vector magnetometer frame to the orientation of the probe laser beam. This could have implications for referencing a vector OPM to non magnetic objects, vector gradiometry calibration, and tracking orientation drifts of the vector OPM not detectable with scalar calibration techniques~\cite{gravrand2001calibration}. 

\begin{backmatter}
\bmsection{Funding} This work was supported by DARPA through ARO grant numbers W911NF-21-1-0127 and W911NF-19-1-0330, NSF QLCI Award OMA - 2016244, and the Baur-SPIE Chair in Optical Physics and Photonics at JILA.

\bmsection{Acknowledgments} We acknowledge helpful conversations with Georg Bison, Michaela Ellmeier, Juniper Pollock, and Dawson Hewatt, and technical expertise from Yolanda Duerst, and Felix Vietmeyer.

\bmsection{Disclosures} The authors declare no conflicts of interest.

\bmsection{Data availability} Data underlying the results presented in this paper are not publicly available at this time but may be obtained from the authors upon reasonable request.

\bmsection{Supplemental document}
See Supplement 1 for supporting content. 

\end{backmatter}


\bibliography{main}

\begin{thebibliography}{10}
\newcommand{\enquote}[1]{``#1''}

\bibitem{kominis2003subfemtotesla}
I.~Kominis, T.~Kornack, J.~Allred, and M.~V. Romalis, \enquote{A subfemtotesla multichannel atomic magnetometer,} {\protect\JournalTitle{Nature}} \textbf{422}, 596--599 (2003).

\bibitem{budker2007optical}
D.~Budker and M.~Romalis, \enquote{Optical magnetometry,} {\protect\JournalTitle{Nature Physics}} \textbf{3}, 227--234 (2007).

\bibitem{dang2010ultrahigh}
H.~Dang, A.~C. Maloof, and M.~V. Romalis, \enquote{Ultrahigh sensitivity magnetic field and magnetization measurements with an atomic magnetometer,} {\protect\JournalTitle{Applied Physics Letters}} \textbf{97} (2010).

\bibitem{sheng2013subfemtotesla}
D.~Sheng, S.~Li, N.~Dural, and M.~V. Romalis, \enquote{Subfemtotesla scalar atomic magnetometry using multipass cells,} {\protect\JournalTitle{Physical Review Letters}} \textbf{110}, 160802 (2013).

\bibitem{bison2003laser}
G.~Bison, R.~Wynands, and A.~Weis, \enquote{A laser-pumped magnetometer for the mapping of human cardiomagnetic fields,} {\protect\JournalTitle{Appl. Phys. B}} \textbf{76}, 325--328 (2003).

\bibitem{xia2006magnetoencephalography}
H.~Xia, A.~Ben-Amar~Baranga, D.~Hoffman, and M.~V. Romalis, \enquote{Magnetoencephalography with an atomic magnetometer,} {\protect\JournalTitle{Appl. Phys. Lett.}} \textbf{89} (2006).

\bibitem{broser2018optically}
P.~J. Broser, S.~Knappe, D.-S. Kajal, \emph{et~al.}, \enquote{{Optically Pumped Magnetometers for Magneto-Myography to Study the Innervation of the Hand},} {\protect\JournalTitle{IEEE Trans. Neural Syst. Rehabil. Eng.}} \textbf{26}, 2226--2230 (2018).

\bibitem{lee2006subfemtotesla}
S.-K. Lee, K.~Sauer, S.~Seltzer, \emph{et~al.}, \enquote{Subfemtotesla radio-frequency atomic magnetometer for detection of nuclear quadrupole resonance,} {\protect\JournalTitle{Applied Physics Letters}} \textbf{89} (2006).

\bibitem{gerginov2017prospects}
V.~Gerginov, F.~Da~Silva, and D.~Howe, \enquote{Prospects for magnetic field communications and location using quantum sensors,} {\protect\JournalTitle{Rev. Sci. Instrum.}} \textbf{88} (2017).

\bibitem{fan2022magnetic}
I.~Fan, S.~Knappe, and V.~Gerginov, \enquote{Magnetic communication by polarization helicity modulation using atomic magnetometers,} {\protect\JournalTitle{Rev. Sci. Instrum.}} \textbf{93} (2022).

\bibitem{lipka2024multiparameter}
M.~Lipka, A.~Sierant, C.~Troullinou, and M.~W. Mitchell, \enquote{Multiparameter quantum sensing and magnetic communication with a hybrid dc and rf optically pumped magnetometer,} {\protect\JournalTitle{Phys. Rev. Appl.}} \textbf{21}, 034054 (2024).

\bibitem{pospelov2013detecting}
M.~Pospelov, S.~Pustelny, M.~P. Ledbetter, \emph{et~al.}, \enquote{{Detecting Domain Walls of Axionlike Models Using Terrestrial Experiments},} {\protect\JournalTitle{Phys. Rev. Lett.}} \textbf{110}, 021803 (2013).

\bibitem{afach2021search}
S.~Afach, B.~C. Buchler, D.~Budker, \emph{et~al.}, \enquote{Search for topological defect dark matter with a global network of optical magnetometers,} {\protect\JournalTitle{Nat. Phys.}} \textbf{17}, 1396--1401 (2021).

\bibitem{pendlebury1984search}
J.~M. Pendlebury, K.~F. Smith, R.~Golub, \emph{et~al.}, \enquote{Search for a neutron electric dipole moment,} {\protect\JournalTitle{Phys. Lett. B}} \textbf{136}, 327--330 (1984).

\bibitem{ayres2021design}
N.~J. Ayres, G.~Ban, L.~Bienstman, \emph{et~al.}, \enquote{The design of the {n2EDM} experiment: {nEDM} {Collaboration},} {\protect\JournalTitle{Eur. Phys. J. C}} \textbf{81}, 512 (2021).

\bibitem{wolfgramm2010squeezed}
F.~Wolfgramm, A.~Cere, F.~A. Beduini, \emph{et~al.}, \enquote{Squeezed-light optical magnetometry,} {\protect\JournalTitle{Phys. Rev. Lett.}} \textbf{105}, 053601 (2010).

\bibitem{troullinou2023quantum}
C.~Troullinou, V.~G. Lucivero, and M.~W. Mitchell, \enquote{Quantum-enhanced magnetometry at optimal number density,} {\protect\JournalTitle{Phys. Rev. Lett.}} \textbf{131}, 133602 (2023).

\bibitem{julsgaard2001experimental}
B.~Julsgaard, A.~Kozhekin, and E.~S. Polzik, \enquote{Experimental long-lived entanglement of two macroscopic objects,} {\protect\JournalTitle{Nature}} \textbf{413}, 400--403 (2001).

\bibitem{kong2020measurement}
J.~Kong, R.~Jim{\'e}nez-Mart{\'\i}nez, C.~Troullinou, \emph{et~al.}, \enquote{Measurement-induced, spatially-extended entanglement in a hot, strongly-interacting atomic system,} {\protect\JournalTitle{Nat. Commun}} \textbf{11}, 2415 (2020).

\bibitem{zhang20224he}
J.~Zhang, Y.~Wang, C.~Wang, \emph{et~al.}, \enquote{{$^4$He Optically Pumped Magnetometer With RF Field Modulation and Light Stabilization in Deep Well for Earthquake Monitoring},} {\protect\JournalTitle{IEEE Transactions on Instrumentation and Measurement}} \textbf{71}, 1--10 (2022).

\bibitem{prouty2016real}
M.~Prouty, \enquote{{Real-Time Hand-Held Magnetometer Array},} Tech. rep., Geometrics (2016).

\bibitem{psiaki1991ground}
M.~L. Psiaki, L.~Huang, and S.~Fox, \enquote{Ground tests of magnetometer-based autonomous navigation {(MAGNAV)} for low-earth-orbiting spacecraft,} {\protect\JournalTitle{Journal of Guidance, Control, and Dynamics}} \textbf{16}, 206--214 (1993).

\bibitem{canciani2016absolute}
A.~Canciani and J.~Raquet, \enquote{Absolute positioning using the {Earth's} magnetic anomaly field,} {\protect\JournalTitle{NAVIGATION: Journal of the Institute of Navigation}} \textbf{63}, 111--126 (2016).

\bibitem{dougherty2004cassini}
M.~Dougherty, S.~Kellock, D.~Southwood, \emph{et~al.}, \enquote{The {Cassini} magnetic field investigation,} {\protect\JournalTitle{Space Science Reviews}} pp. 331--383 (2004).

\bibitem{korth2016miniature}
H.~Korth, K.~Strohbehn, F.~Tejada, \emph{et~al.}, \enquote{Miniature atomic scalar magnetometer for space based on the rubidium isotope {$^{87}$Rb},} {\protect\JournalTitle{Journal of Geophysical Research: Space Physics}} \textbf{121}, 7870--7880 (2016).

\bibitem{bennett2021precision}
J.~S. Bennett, B.~E. Vyhnalek, H.~Greenall, \emph{et~al.}, \enquote{Precision magnetometers for aerospace applications: {A} review,} {\protect\JournalTitle{Sensors}} \textbf{21}, 5568 (2021).

\bibitem{leger2009swarm}
J.-M. Leger, F.~Bertrand, T.~Jager, \emph{et~al.}, \enquote{Swarm absolute scalar and vector magnetometer based on helium 4 optical pumping,} {\protect\JournalTitle{Procedia Chemistry}} \textbf{1}, 634--637 (2009).

\bibitem{alldredge1960proposed}
L.~Alldredge, \enquote{A proposed automatic standard magnetic observatory,} {\protect\JournalTitle{Journal of Geophysical Research}} \textbf{65}, 3777--3786 (1960).

\bibitem{alexandrov2004three}
E.~Alexandrov, M.~Balabas, V.~Kulyasov, \emph{et~al.}, \enquote{Three-component variometer based on a scalar potassium sensor,} {\protect\JournalTitle{Measurement Science and Technology}} \textbf{15}, 918 (2004).

\bibitem{wang2023pulsed}
T.~Wang, W.~Lee, M.~Limes, \emph{et~al.}, \enquote{{High Dynamic Range Vector Atomic Magnetometer with 1 part-per-billion Resolution in Earth Field Range},} {\protect\JournalTitle{arXiv:2304.00214}}  (2024).

\bibitem{gravrand2001calibration}
O.~Gravrand, A.~Khokhlov, J.~L. Le~Mou{\"e}l, and J.~M. L{\'e}ger, \enquote{On the calibration of a vectorial {$^{4}$He} pumped magnetometer,} {\protect\JournalTitle{Earth, Planets and Space}} \textbf{53}, 949--958 (2001).

\bibitem{andryushkov2022vector}
V.~Andryushkov, D.~Radnatarov, and S.~Kobtsev, \enquote{Vector magnetometer based on the effect of coherent population trapping,} {\protect\JournalTitle{Applied Optics}} \textbf{61}, 3604--3608 (2022).

\bibitem{leger2015flight}
J.-M. L{\'e}ger, T.~Jager, F.~Bertrand, \emph{et~al.}, \enquote{In-flight performance of the {Absolute Scalar Magnetometer vector mode on board the Swarm satellites},} {\protect\JournalTitle{Earth, Planets and Space}} \textbf{67}, 1--12 (2015).

\bibitem{zhang2019advanced}
R.~Zhang and R.~Mhaskar, \enquote{{Advanced Magnetometer System: Task II},} Tech. rep., Geometrics (2019).

\bibitem{fairweather1972vector}
A.~Fairweather and M.~Usher, \enquote{A vector rubidium magnetometer,} {\protect\JournalTitle{Journal of Physics E: Scientific Instruments}} \textbf{5}, 986 (1972).

\bibitem{yudin2010vector}
V.~Yudin, A.~Taichenachev, Y.~Dudin, \emph{et~al.}, \enquote{Vector magnetometry based on electromagnetically induced transparency in linearly polarized light,} {\protect\JournalTitle{Physical Review A}} \textbf{82}, 033807 (2010).

\bibitem{gonzalez2024sensitivity}
M.~Gonzalez~Maldonado, O.~Rollins, A.~Toyryla, \emph{et~al.}, \enquote{Sensitivity of a vector atomic magnetometer based on electromagnetically induced transparency,} {\protect\JournalTitle{Optics Express}} \textbf{32}, 25062--25073 (2024).

\bibitem{pustelny2006nonlinear}
S.~Pustelny, W.~Gawlik, S.~Rochester, \emph{et~al.}, \enquote{Nonlinear magneto-optical rotation with modulated light in tilted magnetic fields,} {\protect\JournalTitle{Physical Review A}} \textbf{74}, 063420 (2006).

\bibitem{meng2023machine}
X.~Meng, Y.~Zhang, X.~Zhang, \emph{et~al.}, \enquote{Machine learning assisted vector atomic magnetometry,} {\protect\JournalTitle{Nature Communications}} \textbf{14}, 6105 (2023).

\bibitem{patton2014all}
B.~Patton, E.~Zhivun, D.~Hovde, and D.~Budker, \enquote{All-optical vector atomic magnetometer,} {\protect\JournalTitle{Physical Review Letters}} \textbf{113}, 013001 (2014).

\bibitem{bison2018sensitive}
G.~Bison, V.~Bondar, P.~Schmidt-Wellenburg, \emph{et~al.}, \enquote{Sensitive and stable vector magnetometer for operation in zero and finite fields,} {\protect\JournalTitle{Optics Express}} \textbf{26}, 17350--17359 (2018).

\bibitem{zhang2021vector}
R.~Zhang, R.~Mhaskar, K.~Smith, \emph{et~al.}, \enquote{Vector measurements using all optical scalar atomic magnetometers,} {\protect\JournalTitle{Journal of Applied Physics}} \textbf{129} (2021).

\bibitem{petrenko2023all}
M.~V. Petrenko, A.~S. Pazgalev, and A.~K. Vershovskii, \enquote{All-optical nonzero-field vector magnetic sensor for magnetoencephalography,} {\protect\JournalTitle{Phys. Rev. Appl.}} \textbf{20}, 024001 (2023).

\bibitem{weis2006theory}
A.~Weis, G.~Bison, and A.~S. Pazgalev, \enquote{Theory of double resonance magnetometers based on atomic alignment,} {\protect\JournalTitle{Physical Review A}} \textbf{74}, 033401 (2006).

\bibitem{ingleby2018vector}
S.~J. Ingleby, C.~O’Dwyer, P.~F. Griffin, \emph{et~al.}, \enquote{Vector magnetometry exploiting phase-geometry effects in a double-resonance alignment magnetometer,} {\protect\JournalTitle{Phys. Rev. Appl.}} \textbf{10}, 034035 (2018).

\bibitem{pyragius2019voigt}
T.~Pyragius, H.~M. Florez, and T.~Fernholz, \enquote{Voigt-effect-based three-dimensional vector magnetometer,} {\protect\JournalTitle{Physical Review A}} \textbf{100}, 023416 (2019).

\bibitem{mckelvy2023technical}
J.~McKelvy, M.~Maldonado, I.~Novikova, \emph{et~al.}, \enquote{Technical limits of sensitivity for {EIT} magnetometry,} {\protect\JournalTitle{Applied Optics}} \textbf{62}, 6518--6527 (2023).

\bibitem{horsley2015widefield}
A.~Horsley, G.-X. Du, and P.~Treutlein, \enquote{Widefield microwave imaging in alkali vapor cells with sub-100 $\mu$m resolution,} {\protect\JournalTitle{New journal of physics}} \textbf{17}, 112002 (2015).

\bibitem{horsley2016frequency}
A.~Horsley and P.~Treutlein, \enquote{Frequency-tunable microwave field detection in an atomic vapor cell,} {\protect\JournalTitle{Applied Physics Letters}} \textbf{108} (2016).

\bibitem{bohi2012simple}
P.~B{\"o}hi and P.~Treutlein, \enquote{Simple microwave field imaging technique using hot atomic vapor cells,} {\protect\JournalTitle{Applied physics letters}} \textbf{101} (2012).

\bibitem{nikolova2011microwave}
N.~K. Nikolova, \enquote{Microwave imaging for breast cancer,} {\protect\JournalTitle{IEEE microwave magazine}} \textbf{12}, 78--94 (2011).

\bibitem{chandra2015opportunities}
R.~Chandra, H.~Zhou, I.~Balasingham, and R.~M. Narayanan, \enquote{On the opportunities and challenges in microwave medical sensing and imaging,} {\protect\JournalTitle{IEEE transactions on biomedical engineering}} \textbf{62}, 1667--1682 (2015).

\bibitem{robinson2021determining}
A.~K. Robinson, N.~Prajapati, D.~Senic, \emph{et~al.}, \enquote{Determining the angle-of-arrival of a radio-frequency source with a rydberg atom-based sensor,} {\protect\JournalTitle{Applied Physics Letters}} \textbf{118} (2021).

\bibitem{koepsell2017measuring}
J.~Koepsell, T.~Thiele, J.~Deiglmayr, \emph{et~al.}, \enquote{Measuring the polarization of electromagnetic fields using {Rabi}-rate measurements with spatial resolution: {Experiment} and theory,} {\protect\JournalTitle{Physical Review A}} \textbf{95}, 053860 (2017).

\bibitem{thiele2018self}
T.~Thiele, Y.~Lin, M.~O. Brown, and C.~A. Regal, \enquote{Self-calibrating vector atomic magnetometry through microwave polarization reconstruction,} {\protect\JournalTitle{Physical Review Letters}} \textbf{121}, 153202 (2018).

\bibitem{horsley2013imaging}
A.~Horsley, G.-X. Du, M.~Pellaton, \emph{et~al.}, \enquote{Imaging of relaxation times and microwave field strength in a microfabricated vapor cell,} {\protect\JournalTitle{Physical Review A}} \textbf{88}, 063407 (2013).

\bibitem{kiehl2023coherence}
C.~Kiehl, D.~Wagner, T.-W. Hsu, \emph{et~al.}, \enquote{Coherence of {Rabi} oscillations with spin exchange,} {\protect\JournalTitle{Physical Review Research}} \textbf{5}, L012002 (2023).

\bibitem{olsen2003calibration}
N.~Olsen, L.~T. Clausen, T.~J. Sabaka, \emph{et~al.}, \enquote{Calibration of the {{\O}rsted} vector magnetometer,} {\protect\JournalTitle{Earth, Planets and Space}} \textbf{55}, 11--18 (2003).

\bibitem{kiehl2024correcting}
C.~Kiehl, T.~S. Menon, D.~P. Hewatt, \emph{et~al.}, \enquote{Correcting heading errors in optically pumped magnetometers through microwave interrogation,} {\protect\JournalTitle{Phys. Rev. Appl.}} \textbf{22}, 014005 (2024).

\bibitem{bloom1962principles}
A.~L. Bloom, \enquote{Principles of operation of the rubidium vapor magnetometer,} {\protect\JournalTitle{Applied Optics}} \textbf{1}, 61--68 (1962).

\bibitem{seltzer2008developments}
S.~J. Seltzer, \enquote{Developments in alkali-metal atomic magnetometry,} Ph.D. thesis, Princeton University (2008).

\bibitem{mhaskar2012low}
R.~Mhaskar, S.~Knappe, and J.~Kitching, \enquote{A low-power, high-sensitivity micromachined optical magnetometer,} {\protect\JournalTitle{Applied Physics Letters}} \textbf{101} (2012).

\bibitem{coffer2000atomic}
J.~Coffer and J.~Camparo, \enquote{Atomic stabilization of field intensity using {R}abi resonances,} {\protect\JournalTitle{Phys. Rev. A}} \textbf{62}, 013812 (2000).

\bibitem{kinoshita2010atomic}
M.~Kinoshita, K.~Shimaoka, and K.~Komiyama, \enquote{Atomic {M}icrowave {P}ower {S}tandard {B}ased on the {R}abi {F}requency,} {\protect\JournalTitle{IEEE Trans. Instrum. Meas.}} \textbf{60}, 2696--2701 (2010).

\bibitem{tretiakov2019microwave}
A.~Tretiakov and L.~LeBlanc, \enquote{Microwave {R}abi resonances beyond the small-signal regime,} {\protect\JournalTitle{Phys. Rev. A}} \textbf{99}, 043402 (2019).

\bibitem{mohapatra2007coherent}
A.~Mohapatra, T.~Jackson, and C.~Adams, \enquote{Coherent optical detection of highly excited {Rydberg} states using electromagnetically induced transparency,} {\protect\JournalTitle{Physical Review Letters}} \textbf{98}, 113003 (2007).

\bibitem{sedlacek2012microwave}
J.~A. Sedlacek, A.~Schwettmann, H.~K{\"u}bler, \emph{et~al.}, \enquote{Microwave electrometry with {Rydberg} atoms in a vapour cell using bright atomic resonances,} {\protect\JournalTitle{Nature Physics}} \textbf{8}, 819--824 (2012).

\bibitem{holloway2014broadband}
C.~L. Holloway, J.~A. Gordon, S.~Jefferts, \emph{et~al.}, \enquote{Broadband {Rydberg} atom-based electric-field probe for {SI}-traceable, self-calibrated measurements,} {\protect\JournalTitle{IEEE Transactions on Antennas and Propagation}} \textbf{62}, 6169--6182 (2014).

\bibitem{alexandrov2003recent}
E.~Alexandrov, \enquote{Recent progress in optically pumped magnetometers,} {\protect\JournalTitle{Physica Scripta}} \textbf{2003}, 27 (2003).

\end{thebibliography}

\end{document}


\maketitle

\section{Details and Calibration of the DC coil system}
\begin{figure}[!htbp]
\centering
\includegraphics[width=1\linewidth]{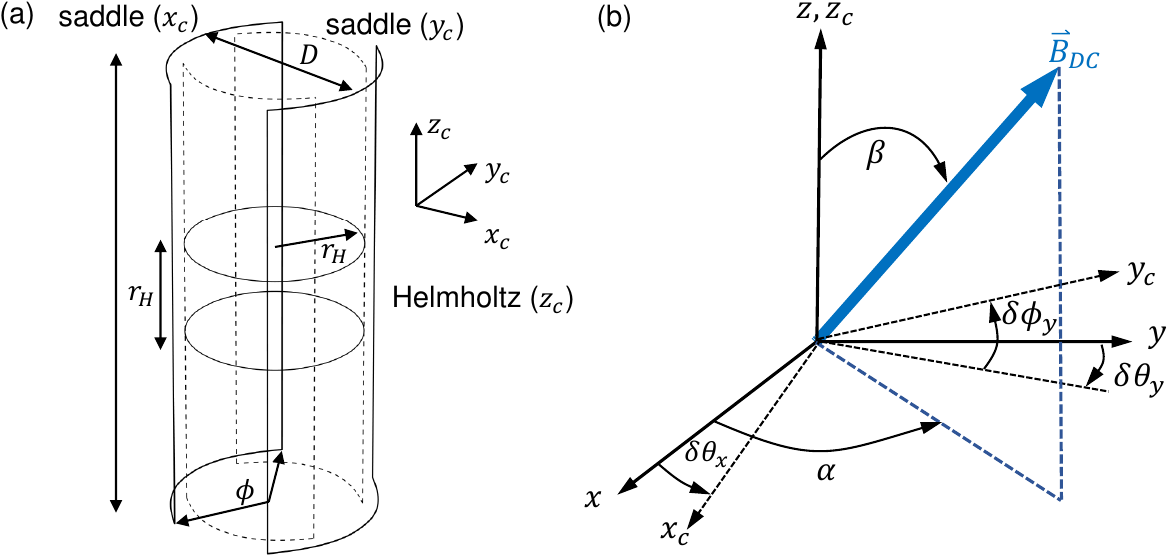}
\caption{(a) Diagram of the 3D coil system consisting of two pairs of saddle coils and a Helmholtz
coil pair. (b) Schematic of the non-orthogonal coil coordinate frame $\mathcal{C}=(x_c,y_c,z_c)$ with respect to the orthogonal laboratory frame $\mathcal{L}=(x, y, z)$.}
\label{fig:coilFrame}
\end{figure}

\begin{figure}[!htb]
\centering
\includegraphics[width=1\textwidth]{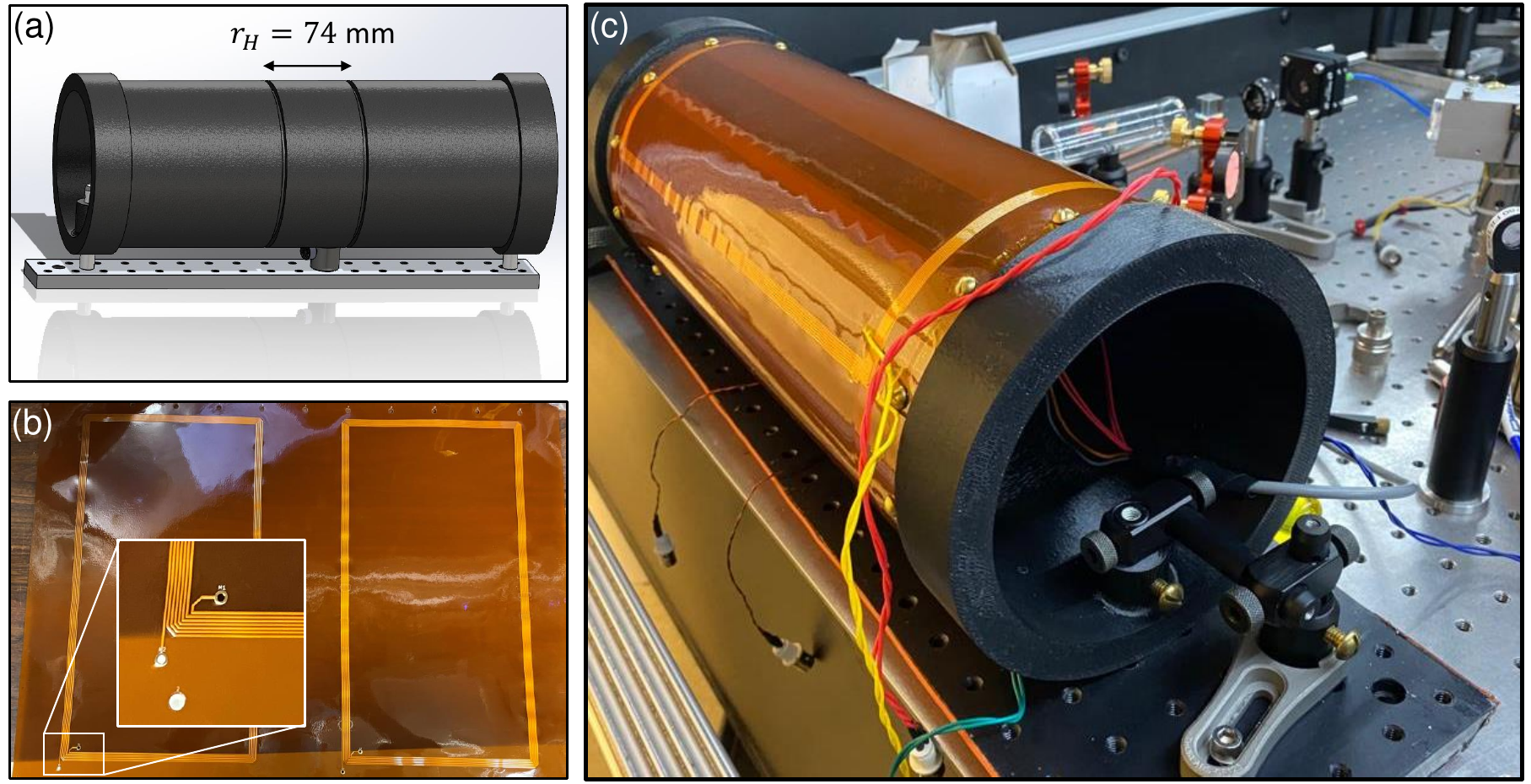}
\caption{The 3D coil system. (a) A CAD drawing of the cylindrical coil mount. Groves are made to wind the Helmholtz coils. (b) The flexible pcb saddle coil. (c) The coil system put together. The mounting cylinder (black) is 3D printed of nylon with the flexible pcb saddle coils wrapped around.}
\label{fig:coilSystemPhysical}
\end{figure}

The DC coil system consists of three near-orthogonal coil pairs that generate fields along coil directions $(\hat{x}_c,\hat{y}_c,\hat{z}_c)$ given in terms of coil currents $(I_x,I_y,I_z)$
\begin{align}
\vec{B}_{x,c}=I_xa_x(1+\epsilon_x)\vec{x}_c\\
\vec{B}_{y,c}=I_ya_y(1+\epsilon_y)\vec{y}_c\\
\vec{B}_{z,c}=I_za_z(1+\epsilon_z)\vec{z}_c.
\end{align}
Here $(a_x,a_y,a_z)=(91.6926,91.2159,392.773)$ $\mu$T/A are pre-calibrated coil coefficents and $(\epsilon_x,\epsilon_y,\epsilon_z)$ are coil correction terms to be determined. The control of the $B_{x,c}$ and $B_{y,c}$ magnetic fields are managed by saddle coils, characterized by diameter $D=148$ mm, height $h=300$ mm, and central angle subtended by the saddle coils $\phi=120^{\circ}$ [Fig.~\ref{fig:coilFrame}(a)]. The $B_{z,c}$ control is managed by a Helmholtz coil pair characterized by radius $r_H=74$ mm. To implement this coil system, we used a large 3D printed nylon cylindrical structure with $4\times5$ mm$^2$ groves made for the Helmholtz pair [Fig.~\ref{fig:coilSystemPhysical}(a)], using approximately 30 turns of wire. The saddle coils were constructed from a flexible six-layer PCB, featuring seven turns per layer and trace widths measuring 0.5 mm [Fig.~\ref{fig:coilSystemPhysical}(b,c)]. For the few hundred mA of current used in the Helmholtz pair and saddle coils to produce magnetic fields near $50$ $\mu$T, we observed coil heating limited to within 3 \textdegree C. 

We establish an orthogonal laboratory frame $\mathcal{L}=(x,y,z)$ with respect to the DC coil system frame $\mathcal{C}=(x_c,y_c,z_c)$ as follows:
\begin{align}
\hat{x}_c=R_y(\pi/2+\delta \theta_x)\hat{z}=\{\text{cos}[\delta \theta_x],0,-\text{sin}[\delta\theta_x]\}\\
\hat{y}_c=R_z(\pi/2+\delta \phi_y)R_y(\pi/2+\delta \theta_y)\hat{z}=\{-\text{cos}[\delta \theta_y]\text{sin}[\delta\phi_y],\text{cos}[\delta \theta_y]\text{cos}[\delta \phi_y],-\text{sin}[\delta \theta_y]\}\\
\hat{z}_c=\hat{z}=\{0,0,1\}
\end{align}
where $R_{y,z}$ are rotation matrices about the $\hat{y}$ and $\hat{z}$ directions, and $(\delta \theta_x, \delta \theta_y, \delta \phi_y)$ are non-orthogonality angles. The total field $|\vec{B}|$ generated by the coil system and a background field $\vec{B}_0=(B_{x,o},B_{y,o},B_{z,o})$ is given by
\begin{equation}
|\vec{B}|^2=\Big(B_{x,o}+\sum_{k=x,y,z}\vec{B}_{k,c}\cdot\hat{x}\Big)^2+\Big(B_{y,o}+\sum_{k=x,y,z}\vec{B}_{k,c}\cdot\hat{y}\Big)^2+\Big(B_{z,o}+\sum_{k=x,y,z}\vec{B}_{k,c}\cdot\hat{z}\Big)^2.
\end{equation}
In this framework there are nine unknown parameters: three non-orthogonality angles $(\delta \theta_x, \delta \theta_y, \delta \phi_y)=(3.72,-0.99,3.47)$ mrad, three coil corrections $(\epsilon_x, \epsilon_y, \epsilon_z)=(0.57,0.16,2.7)\times10^{-3}$, and three background field components $(B_{x,o},B_{y,o},B_{z,o})=(-84.0,50.7,-62.2)$ nT. We fit these nine parameters using scalar measurements ($B$) by minimizing
\begin{equation}
C_{\text{coil}}=\sum_{j=1}^{60}w_j\big(|\vec{B}|_j-B_j\big)^2
\end{equation}
where $j$ denotes a programmed current 3-tuple $(I_x,I_y,I_z)_j$ that corresponds to a magnetic field direction $(\alpha,\beta)_j$ with norm $B_j\approx 50$ $\mu$T. The DC magnetic field strengths are fitted from FID measurements using a double-frequency model accounting for the different precessional frequencies in the $F=1,2$ manifolds~\cite{kiehl2024correcting}. The weights $w_j=1/\delta B ^2$ are given by the variance $\delta B ^2$ calculated from 10 repeated FID measurements.

To account for coil system drift during the 37.5-minute vector data acquisition, we perform running calibrations of coil parameters post-processing using FID measurements taken at various magnetic field directions from the $(\alpha,\beta)$ sequence, which includes both random and repeated field directions, as described in Fig. 2(b) and Fig. 3(a) of the main text. Specifically, we perform coil system calibrations on consecutive chunks of 120 field directions from this sequence, each lasting 5 minutes. However, FID measurements from only 60 of the 120 directions are used for the calibration. The 5-minute duration was chosen to ensure minimal coil system drift during calibration, meaning the system remains stable over this time frame. As described below, the directional accuracy of the magnetic field programmed by our coil system is within 50 $\mu$rad.


\begin{figure}[!htbp]
\centering
\includegraphics[width=1\linewidth]{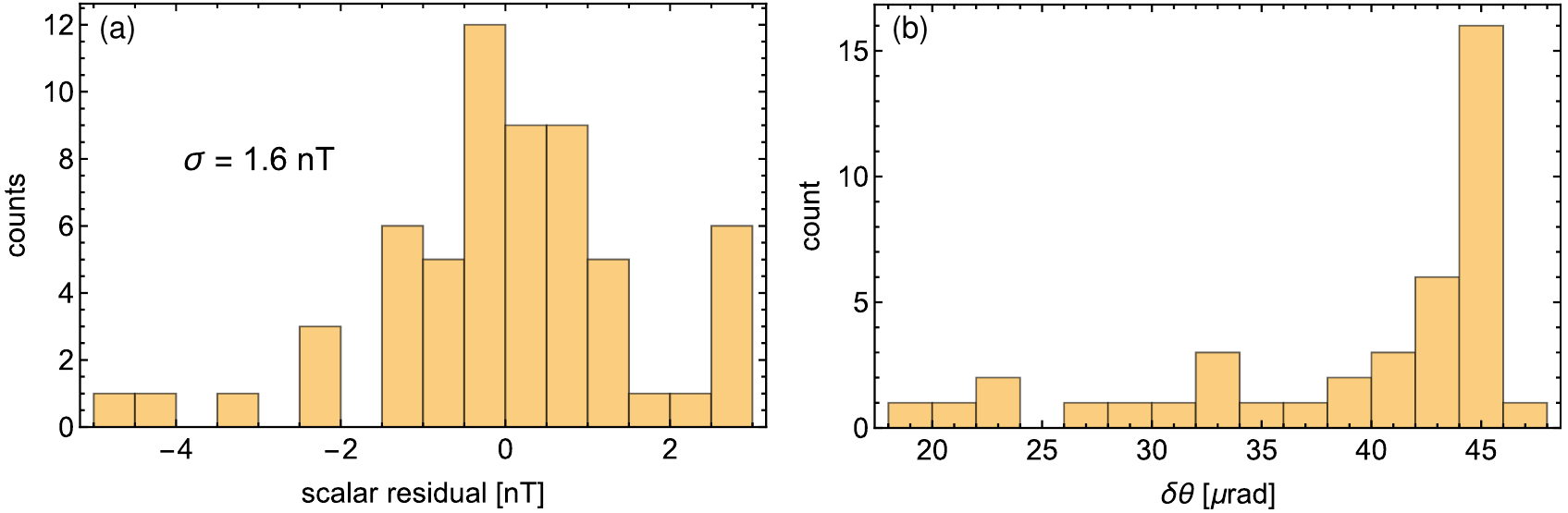}
\caption{Coil system calibration residuals and estimated directional accuracy. (a) Calibration residuals using FID measurements evaluated over 60 random DC magnetic field direction. (b) Estimated errors in the direction of $\vec{B}$ assuming 2.3 nT error in $B_{x,o}$. A similar histogram is obtained for estimated $\delta \theta$ errors assuming an error of $\delta \epsilon_x=0.06\times10^{-3}$ for $\epsilon_x$ bounded by 50 $\mu$rad.}
\label{fig:coilAcc}
\end{figure}

The residuals of one of these calibrations are shown in Fig.~\ref{fig:coilAcc}(a). The calibration residuals of a few nT, with a standard deviation of 1.6 nT, remain consistent across all of the calibrations. These residuals are substantially larger than the average statistical standard error, $\delta B <100$ pT, derived from 10 repeated FID measurements, indicating the presence of systematic errors. Two likely sources of these systematic errors are heading error in the FID measurement~\cite{kiehl2024correcting} and nonlinearities in the coil current control.

To assess the accuracy of the programmed field direction within the lab frame $\mathcal{L}$, we intentionally alter one of the coil parameters in the system model (e.g., $\epsilon_x \rightarrow \epsilon_x + \delta \epsilon_x$ or $B_{x,0} \rightarrow B_{x,0} + \delta B_{x,0}$) and then compare the resulting $|\vec{B}|$ to that of the unperturbed coil model. The parameters $\epsilon_x$ and $B_{x,0}$ were chosen as they represent two distinct potential drift scenarios in the coil system: a change in the coil factors or a shift in the background magnetic field. To produce a root-mean-square deviation (RMSD) of 1.6  nT between scalar measurements predicted by the unperturbed and perturbed coil models, which matched the standard deviation of the measured calibration residuals, required $\delta B_{x,0}=2.3$ nT or $\delta \epsilon_x=0.06\times10^{-3}$ in the perturbed coil model. The deviations in the directions of the DC magnetic fields between the unperturbed and perturbed coil models are shown in [Fig.~\ref{fig:coilAcc}(b)]. In both cases these directional deviations are confined within 50 $\mu$rad. Consequently, we estimate that the directional accuracy of our coil system throughout the vector measurements is maintained within 50 $\mu$rad.

\section{Vapor cell temperature and buffer gas pressure monitoring}

\begin{figure}[!htbp]
\centering
\includegraphics[width=1\linewidth]{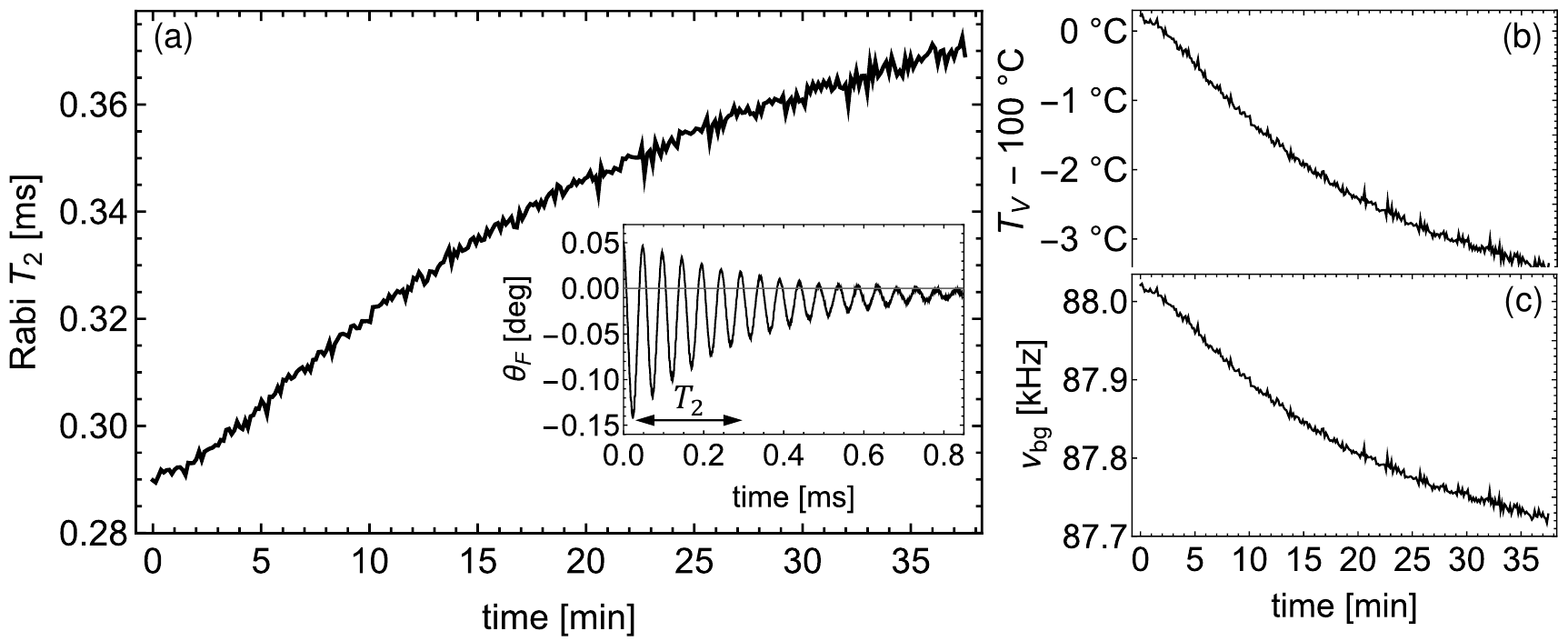}
\caption{Drift monitoring of the vapor temperature ($T_v$) and buffer gas pressure shift ($\nu_{\text{bg}}$) through the Rabi dephasing time $T_2$ . (a) Rabi dephasing time over the duration of the vector measurements. These measurements use MPE 1 $\sigma^+$ Rabi oscillations at the magnetic field direction $(135\degree,57\degree)$. (b) and (c) show the predicted vapor temperature and buffer gas pressure shift respectively.}
\label{fig:pressureShiftDrift}
\end{figure}

This section discusses how to monitor the vapor temperature ($T_{\text{v}}$) and the corresponding changes in the buffer gas pressure ($\nu_{\text{bg}}$) through the Rabi oscillation dephasing time $T_2$ [Fig.~\ref{fig:pressureShiftDrift}]. This monitoring is important because changes in the buffer gas pressure will lead to changes in the microwave detuning during Rabi measurements. For this monitoring, we use the MPE 1 $\sigma^+$ data taken at the repeated magnetic field orientation $(\alpha,\beta)=(135\degree,57\degree)$. 

The primary reason for the microwave field drift depicted in Fig.~3(d) of the main text is cooling of the microwave cavity by a few degrees Celsius over 37.5 minutes. This cooling arises from the fact that the cavity heaters are turned off during the atomic measurements.
The buffer gas pressure shift at the start of these measurements is initially assumed to be $\nu_{\text{bg}}=88$ kHz from independent pressure shift characterizations discussed in Ref.~\cite{kiehl2024correcting}. In this work relative changes in $\nu_{\text{bg}}$ due to drifts in the vapor temperature $T_{\text{v}}$ [Fig.~\ref{fig:pressureShiftDrift}(b)] are estimated from the Rabi oscillation dephasing rate $1/T_2 \approx n_{\text{Rb}}\sigma_{\text{se}}v^r$,
where $n_{\text{Rb}}=P_{\text{Rb}}(T_{\text{v}})/k_B T_{\text{v}}$ is the atomic density expressed in terms of the Rb vapor pressure $P_{\text{Rb}}(T_{\text{v}})$ and the Boltzmann constant $k_B$, $\sigma_{\text{se}}=1.9\times10^{-18}$ cm$^2$ is the Rb spin-exchange (SE) collisonal cross-section~\cite{gibbs1967spin}, and $v^r=\sqrt{8 k_B T_{\text{v}}/\pi m_r}$ is the mean speed between colliding Rb atoms with reduced mass $m_r$. As an approximation, we focus solely on dephasing from spin-exchange (SE) collisions in our temperature estimation, as it represents the predominant mechanism of decoherence~\cite{kiehl2023coherence}. We estimate changes in $\nu_{\text{bg}}$ from drifts in $T_{\text{v}}$ through~\cite{vanier1989quantum}
\begin{equation}
\frac{\nu_{\text{bg}}}{\nu_{0}}=P_0[\beta_0+\delta_0(T_{\text{v}}-T_{\text{v},0})]
\end{equation}
where $\nu_{0}=6834.682$ MHz is the $5^2$S$_{1/2}$ hyperfine splitting for $^{87}$Rb, $T_{\text{v},0}=60$ $\degree$C, $P_0=151.39$ Torr (buffer gas pressure at $T_{\text{v,0}}$), $\beta_0/\nu_{0}=81.9\times 10^{-9}$ Torr$^{-1}$, and $\delta_0/\nu_{0}=79\times 10^{-12}$ $\degree$C$^{-1}$Torr$^{-1}$. The estimated drift in the buffer gas pressure shift from this analysis presented in Fig.~\ref{fig:pressureShiftDrift}(c) is incorporated within the generalized Rabi frequency model $\delta \lambda_{m}^{m^{\prime}}$ when fitting to the Rabi measurements. 

While our estimations of the drifts of $\nu_{\text{bg}}$ from the Rabi dephasing time are only approximate due to our simplified model for $T_2$, the generalized Rabi frequency $(\tilde{\Omega}_m^{m^{\prime}})_p$ is only sensitive to changes in the microwave detuning to second-order because the Rabi oscillations are driven near resonance. Therefore, errors in estimating $\nu_{\text{bg}}$ by about 100 Hz would correspond to only minor modeling errors in $(\tilde{\Omega}_m^{m^{\prime}})_p$, estimated within 1 Hz. Thus, although we use an indirect method to monitor $\nu_{\text{bg}}$ drifts, we do not expect errors in the tracking of $\nu_{\text{bg}}$ to be a dominant systematic error for the vector measurements.

\section{Additional details on the SP\MakeLowercase{a}R fitting algorithm}
This section covers additional details behind the SPaR fitting algorithm including the process for eliminating Larmor frequency components from the SPaR spectra, how to linearize the eigenvalues of the atom-microwave Hamiltonian, and details behind fitting to the $\pi$ transitions. The SPaR fitting algorithm for the $\sigma^{\pm}$ transitions is fully covered in Sec.~6 of the main text.
\subsection{Removal of Larmor frequency components}

\begin{figure}[htbp]
\centering
\includegraphics[width=1\linewidth]{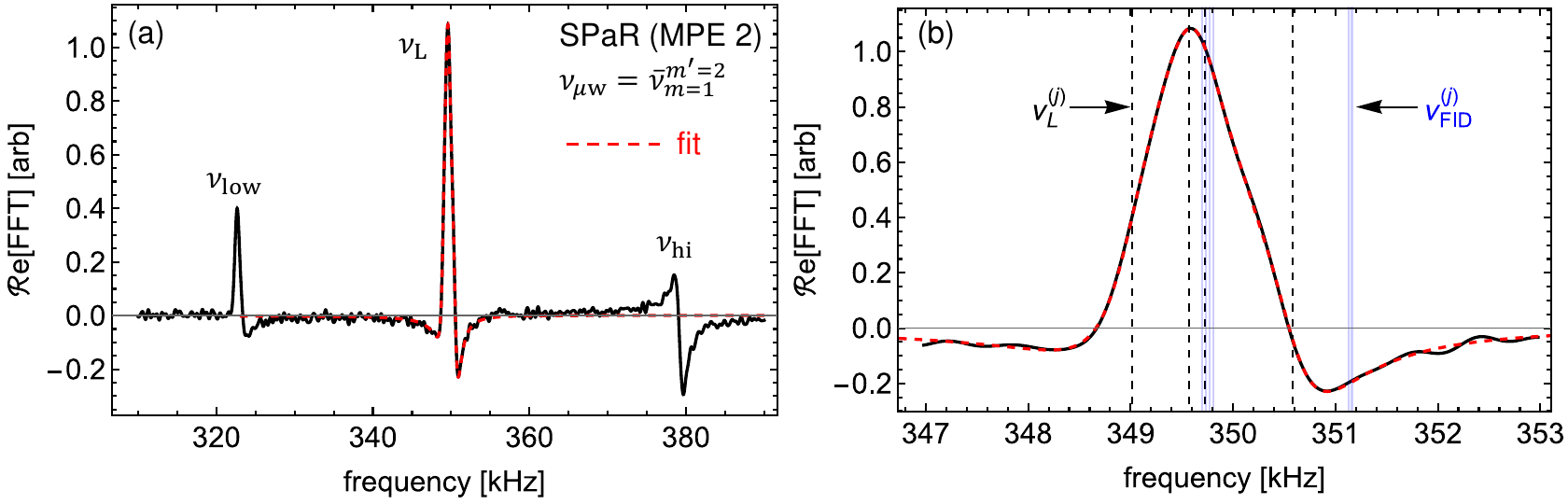}
\caption{Fitting the $\nu_L$ frequency components within SPaR spectra. (a) MPE 2 $\sigma^+$ SPaR spectra with the $\nu_L$ fit overlaid. (b) A close-up view of the $\nu_L$ feature composed of four frequency components $\nu_L^{(j)}$ (black dashed lines), and calculated from the eigenvalues of $H$ using initial Rabi rates estimated from the $\nu_{\text{low/hi}}$ SPaR peaks. Additionally, the unperturbed frequency components of the Larmor FID signal are shown (light blue).}
\label{fig:larmorSub}
\end{figure}

As mentioned in Sec.~6 of the main text, we remove the broadened central Larmor $\nu_L$ peak to achieve consistent SPaR fitting across different DC magnetic field and microwave field configurations. The broadened central Larmor $\nu_L$ peak consists of multiple frequency components that are marked as dashed vertical lines in Fig.~\ref{fig:larmorSub}(b). These $\nu_L$ frequency components are each calculated from eigenvalue pairs of the atom-microwave Hamiltonian $H$ defined in Eq.~4 of the main text. If the microwave field is turned off ($\Omega_m^{m^{\prime}}=0$), then the $\nu_{L}$ components reduce to the six frequency components ($\nu_{\text{FID}}^{(j)}$), marked by light blue vertical lines in Fig.~\ref{fig:larmorSub}(b), corresponding to nonlinear Zeeman (NLZ) shifts originating from the $F=1,2$ hyperfine
manifolds~\cite{kiehl2024correcting}. In contrast, when applying near-resonant microwave coupling there are consistently four components ($\nu_L^{(j)}$).

\begin{figure}[htbp]
\centering
\includegraphics[width=1\linewidth]{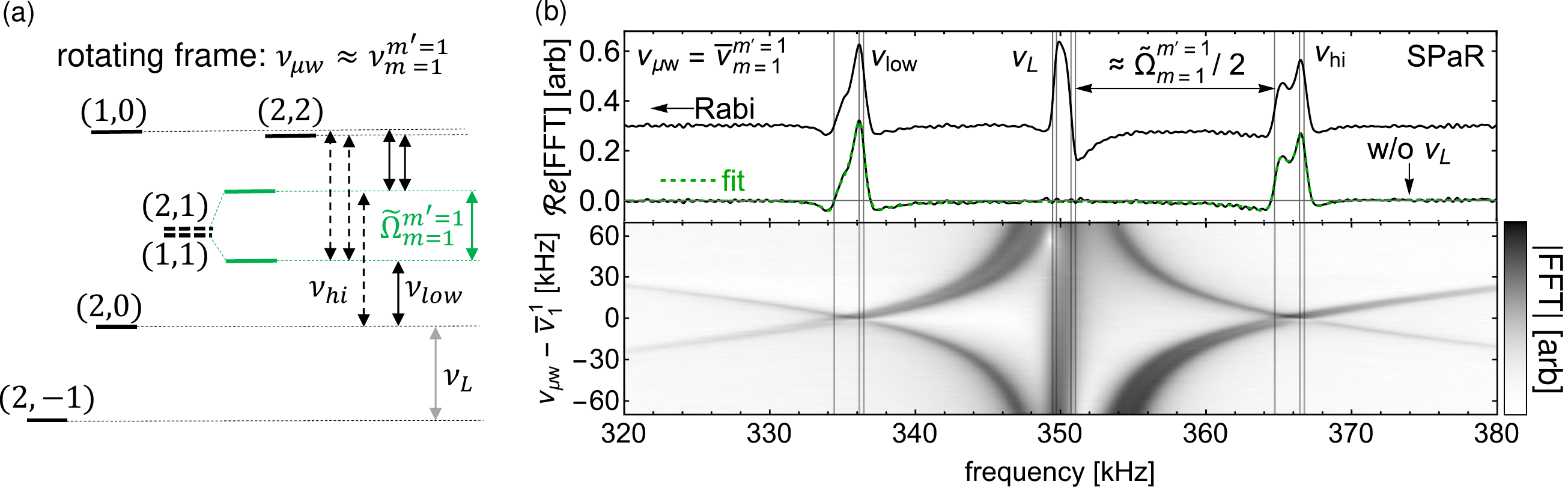}
\caption{SPaR for the $\ket{11}-\ket{21}$ $\pi$ transition. (a) Level diagram in a rotating frame near the $\pi$ hyperfine transition resonance $\nu_{m=1}^{m^{\prime}=1}$. The anti-crossing (green) gives rise to several SPaR peaks above ($\nu_{\text{hi}}$ denoted as dashed arrows) and below ($\nu_{\text{low}}$ denoted as solid arrows) the central Larmor frequency peak ($\nu_L$). The Larmor frequency peak $\nu_L$ arises from the Zeeman splitting between sublevels weakly perturbed by the microwave field. (b) The SPaR spectrum evaluated at $\overline{\nu}_{m=1}^{m^{\prime}=1}=6835.4701$ MHz. Solid vertical lines indicate eigenvalue pairs that correspond to arrows in (a). The bottom density plot shows the microwave detuning dependence of the SPaR spectrum.} 
\label{fig:piSPARS}
\end{figure}

To calculate the four $\nu_L^{(j)}$ components requires estimates of the spherical microwave components, which can be obtained from the Rabi rates $\Omega_m^{m^{\prime}}$ through Eq.~1 of the main text. 
The Rabi rates $|\Omega_m^{m^{\prime}}|\approx \nu_{\text{hi}}-\nu_{\text{low}}$ are estimated from the $\nu_{\text{low/hi}}$ peaks in the fast Fourier transform (FFT) of the $\sigma^{\pm}$ SPaR spectra [Fig.~\ref{fig:larmorSub}(a)] with an accuracy of 1 kHz. These Rabi rates are approximated to be real and positive. The 1 kHz accuracy of these initial Rabi rates affects the accuracy of computing the $\nu_L$ frequency components at the 10 Hz scale. We fix the $\nu_L^{(j)}$ frequency component estimates within the following model for the $\mathcal{R}e[\text{FFT}]$ of the SPaR spectra
\begin{equation}
\label{eq:FSP}
\mathcal{R}e[\text{FFT}] = \sum_{j=1}^4a_j\frac{\text{cos}[\phi_j]-\text{sin}[\phi_j](f-\nu_L^{(j)})}{(f-\nu_L^{(j)})^2+w_j^2/4}
\end{equation}
where the free fitting parameters $a_j$, $w_j$, and $\phi_j$ correspond to Lorentzian amplitudes, widths, and phases. We fit this model to the $\mathcal{R}e[\text{FFT}]$ data over $\nu_{\text{FID}}\pm 3$ kHz, where $\nu_{\text{FID}}\approx350$ kHz is FID frequency when the microwave field is off ($\Omega_m^{m^{\prime}}=0$), and subtract it from the SPaR spectra as shown for the $\pi$ $\ket{11}-\ket{21}$ transition in Fig.~\ref{fig:piSPARS}(b) and the $\sigma^+$ $\ket{11}-\ket{22}$ transition in Fig.~4(d) of the main text.

\subsection{Linearization}
Fitting SPaR spectra by constantly reevaluating the eigenvalues $\lambda$ of the atom-microwave Hamiltonian $H$ is computationally intensive and slow. To circumvent this process, we linearize the eigenvalues about spherical microwave component estimates $(\overline{\mathcal{B}}_{\sigma^+}$, $\overline{\mathcal{B}}_{\pi}$, $\overline{\mathcal{B}}_{\sigma^-})$ with 100 nT accuracy using a peak-finding algorithm of the $\sigma^{\pm}$ SPaR spectra and the calibrated microwave amplitude $|\vec{\mathcal{B}}|$ discussed in Sec.~6 of the main text. For simplicity, we omit the ${(\alpha,\beta)}$ notation from the usual $\mathcal{B}_k^{(\alpha,\beta)}$ in this section. During SPaR fitting, we approximate all spherical microwave components and Rabi rates as real and positive. The Rabi rates are derived from the spherical microwave components using Eq.~1 of the main text.

The linearization of any given eigenvalue $\lambda_j$ takes the form
\begin{equation}
\lambda_{j}\Big\vert_{\mathcal{B}_{\sigma^+},\mathcal{B}_{\pi},\mathcal{B}_{\sigma^-}}=\lambda_{j}\Big\vert_{\overline{\mathcal{B}}_{\sigma^+},\overline{\mathcal{B}}_{\pi},\overline{\mathcal{B}}_{\sigma^-}}+\gamma_+(\mathcal{B}_{\sigma^+}-\overline{\mathcal{B}}_{\sigma^+})+\gamma_{\pi}(\mathcal{B}_{\pi}-\overline{\mathcal{B}}_{\pi})+\gamma_-(\mathcal{B}_{\sigma^-}-\overline{\mathcal{B}}_{\sigma^-})
\end{equation}
where 
\begin{equation}
\gamma_+=\frac{1}{\delta}\Big(\lambda_{j}\Big\vert_{\overline{\mathcal{B}}_{\sigma^+}+\delta,\overline{\mathcal{B}}_{\pi},\overline{\mathcal{B}}_{\sigma^-}}-\lambda_{j}\Big\vert_{\overline{\mathcal{B}}_{\sigma^+},\overline{\mathcal{B}}_{\pi},\overline{\mathcal{B}}_{\sigma^-}} \Big)
\end{equation}
with similar formulas for $\gamma_{\pi}$ and $\gamma_-$. We use $\delta=100$ nT for calculating $\gamma_{\pm}$ and $\gamma_{\pi}$. Systematic errors in the eigenvalue linearization are on the order of $h\cdot$(100 Hz) for 100 nT errors in initial spherical microwave component estimates that are dominated by neglecting second order cross-terms (e.g. terms $\propto (\mathcal{B}_{\pi}-\overline{\mathcal{B}}_{\pi})\times(\mathcal{B}_{\sigma^{+}}-\overline{\mathcal{B}}_{\sigma^{+}})$). These errors could be further compensated by generalizing the linearizaton to second-order about the initial spherical microwave component estimates.

\subsection{SPaR for $\pi$ transitions}
In Fig.~\ref{fig:piSPARS} we show the SPaR spectrum for the $\ket{1,1}-\ket{2,1}$ $\pi$ transition. A similar spectrum appears for the other $\ket{1,-1}-\ket{2,-1}$ $\pi$ transition. This spectrum contains more complexity than the $\sigma^{\pm}$ cases due to the fact that these $\pi$ hyperfine transitions are not end-resonances involving sublevels with maximum or minimum atomic spin projection. We can qualitatively identify the frequency components of the $\nu_{\text{low/hi}}$ sidebands in the level diagram of Fig.~\ref{fig:piSPARS}(a). The fitting model used for $\pi$ SPaR spectra is given by 
\begin{align}
\begin{split}
\label{eq:piSPARS}
\mathcal{R}e[\text{FFT}] =& \sum_{j=1}^3 a_{j,\text{low}}\frac{\text{cos}[\phi_{j,\text{low}}]-\text{sin}[\phi_{j,\text{low}}](f-\nu_{\text{low}}^{(j)})}{(f-\nu_{\text{low}}^{(j)})^2+w_{j,\text{low}}^2/4} \\ &+\sum_{j=1}^3 a_{j,\text{hi}}\frac{\text{cos}[\phi_{j,\text{hi}}]-\text{sin}[\phi_{j,\text{hi}}](f-\nu_{\text{hi}}^{(j)})}{(f-\nu_{\text{hi}}^{(j)})^2+w_{j,\text{hi}}^2/4}
\end{split}
\end{align}
where $a_{j,\text{low(hi)}}$, $w_{j,\text{low(hi)}}$, and $\phi_{j,\text{low(hi)}}$ are free Lorentzian amplitude, widths, and phases for the three $\nu_{\text{low(hi)}}$ frequency components. Only the $\mathcal{B}_{\pi}$ spherical microwave component is left as a free variable to evaluate the $\nu_{\text{low(hi)}}$ frequency components (given in terms of the eigenvalues of $H$), while the $\mathcal{B}_{\sigma^{\pm}}$ components are held at their initial estimates $\overline{\mathcal{B}}_{\sigma^+}$ and $\overline{\mathcal{B}}_{\sigma^-}$. Due to the larger number of free parameters in ~\eqref{eq:piSPARS} the $\pi$ SPaR fitting algorithm is more susceptible to systematic errors from overfitting than the $\sigma^{\pm}$ SPaR model discussed in Sec.~6 of the main text.

\section{Expected Rabi modeling errors due to finite MPE calibration length and measurement noise}
\label{sec:rabiModelErrors}

In this section, given the noise of generalized Rabi frequency measurements, we estimate the size of generalized Rabi frequency residuals against the eigenvalue model $\delta \lambda_m^{m^{\prime}}$ during MPE calibration. We define the Rabi frequency residual as
\begin{equation}
\label{eq:rabRes}
(\delta_m^{m^{\prime}})_p =(\tilde{\Omega}_m^{m^{\prime}})_p-\delta \lambda_m^{m^{\prime}}((\vec{\mathcal{B}}_m^{m^{\prime}})_p,\alpha,\beta).
\end{equation} In Fig.~3(c) of the main text, these residuals are evaluated for the $\sigma^+$ MPE 2 measurements, showing a standard deviation of $\sigma = 13$ Hz. Therefore, for simplicity, we concentrate this analysis on the $\sigma^+$ MPE 2 data.

Our approach is to generate fake MPE 2 Rabi data with the same MPE parameters and DC magnetic field orientations used in the experiment, along with added noise characterized by the known variance of repeated Rabi oscillation measurements. We then calibrate MPE 2 parameters from this fake data and examine the calibration residuals. If the standard deviation of the calibration residuals with the fake data is lower than $\sigma = 13$ Hz, this indicates the presence of systematic errors in the Rabi measurements. From the Rabi measurement noise and these systematic error estimates, we also estimate the uncertainty of the MPE parameters given the finite calibration length $N=12$. For notational simplicity, we use $(\tilde{\Omega}_{m=1}^{m^{\prime}=2})_{s=2} \rightarrow \Omega_{\sigma^+}$, with analogous notation for the microwave phasor, transition dipole moments, and other related terms. 

To begin, we first reiterate from Sec.~4 of the main text that the MPE phasor $\vec{\mathcal{B}}$ associated with each Rabi measurement at a randomly chosen angle $(\alpha,\beta)$ is determined using two MPE phasors, $\vec{\mathcal{B}}^{(1)}$ and $\vec{\mathcal{B}}^{(2)}$. Phasors $\vec{\mathcal{B}}^{(1)}$ and $\vec{\mathcal{B}}^{(2)}$ are each calibrated with generalized Rabi frequencies measured at 12 DC magnetic field orientations different from the choosen $(\alpha,\beta)$ to avoid overfitting. For the analysis in this section, we neglect $\vec{\mathcal{B}}^{(2)}$ and use the measured $\vec{\mathcal{B}}^{(1)}$ to generate fake $\sigma^+$ Rabi data corresponding to all $12+12=24$ Rabi frequencies corresponding to the 24 DC magnetic field directions. We then generate fake Rabi data using this procedure for the various $\vec{\mathcal{B}}^{(1)}$ values corresponding to the ongoing calibrations performed in the experiment across different magnetic field directions.

We also simplify our analysis by using Rabi rates instead of generalized Rabi frequencies. This fake Rabi rate data ($\Omega_{\sigma^+,G}$), with added Gaussian noise characterized by variance $(\delta \Omega_{\sigma^+})^2$, is given by
\begin{equation}
\label{eq:RabiWithNoise}
\Omega_{\sigma^+,G} = G \Big( |\mu_{\sigma^+} \mathcal{B}_{\sigma^+}^{(\alpha,\beta)}/h| ,(\delta \Omega_{\sigma^+})^2 \Big)
\end{equation}
where
\begin{equation}
\label{eq:suppPlusComp}
\mathcal{B}_{\sigma^+}^{(\alpha,\beta)}=R_y(-\beta)R_z(-\alpha)\vec{\mathcal{B}}^{(1)}\cdot\epsilon_+
\end{equation}
is the $\sigma^+$ microwave component calculated from $\vec{\mathcal{B}}^{(1)}$ parameters with $\epsilon_+= \{\frac{1}{\sqrt{2}},- \frac{i}{\sqrt{2}},0\}$. The notation $G(\mu,\sigma^2)$ represents a normal distribution with mean $\mu$ and variance $\sigma^2$. Here, $\delta \Omega_{\sigma^+}$ is the corresponding standard error calculated from 10 repeated Rabi measurements. 

With this set of fake Rabi data, we recalibrate each $\vec{\mathcal{B}}^{(1)}\rightarrow \vec{\mathcal{B}}^{(1')}$, using the associated $N=24$ DC magnetic field directions. Each new phasor $\vec{\mathcal{B}}^{(1^{\prime})}$ is not equal to the original (true) $ \vec{\mathcal{B}}^{(1)}$ due to the added noise and finite calibration length. We then evaluate the calibration residuals through
\begin{equation}
\label{eq:resFromNoieAndMPE}
\delta_{\sigma^+,G}=\Omega_{\sigma^+,G}-\frac{\mu_{\sigma^+}}{h}R_y(-\beta)R_z(-\alpha)\vec{\mathcal{B}}^{(1^{\prime})}\cdot\epsilon_+.
\end{equation}

From this analysis we find that the expected standard deviation of the residuals \eqref{eq:rabRes} evaluated over 300 random field directions is $\sigma=5$ Hz due to measurement noise and finite MPE calibration length. In comparison, $\sigma=3.4$ Hz due solely to the finite MPE calibration length by evaluating
\begin{equation}
\label{eq:resFromMPEAlone}
\delta_{\sigma^+}=\frac{\mu_{\sigma^+}}{h}(|R_y(-\beta)R_z(-\alpha)\vec{\mathcal{B}}^{(1^{\prime})}\cdot\epsilon_+|-|R_y(-\beta)R_z(-\alpha)\vec{\mathcal{B}}^{(1)}\cdot\epsilon_+|).
\end{equation}
This $\sigma=3.4$ Hz represents systematic errors from finite MPE calibration length alone. We summarize these results in Table~\ref{tab:SystematicErrorRes}. 

The fact that the predicted residuals ($\sigma=5$ Hz) are less than the measured residuals ($\sigma=13$ Hz), as reported in Sec.~4 of the main text, implies that the Rabi measurements contain systematic errors beyond that due to finite MPE calibration length. By incorporating larger noise in the simulated Rabi rate data (i.e. larger $\delta \Omega_{\sigma^+}$) so that the standard deviation of the predicted calibration residuals match with the observed value of $\sigma=13$ Hz, and analyzing how the distribution of calibrated MPE parameters deviates from the distribution of calibrated MPE parameters when using the true experimental $\delta \Omega_{\sigma^+}$, we can gauge the magnitude of systematic errors present in the calibrated MPE parameters. This analysis predicts systematic errors of $\delta \mathcal{B}_x=0.4$ nT, $\delta\mathcal{B}_y=0.3$ nT, $\delta\mathcal{B}_z=0.2$ nT, $\delta (\phi_y-\phi_x)=0.3$ mrad as deduced from the standard deviation of repeated calibrations. We explore potential systematic error sources in Sec. \ref{sec:systematicSources}.

\begin{table}[h]
\caption{\label{tab:SystematicErrorRes}
Comparisons of the Rabi frequency fit residuals using simulated and measured data evaluated across random DC magnetic field orientations. If the systematic errors in the vector measurements are solely due to the finite MPE calibration length, the standard deviations in the second and third rows would be identical.}
\begin{center}
\begin{tabular}{|c||c|c|}
\hline
\textrm{Type of Rabi frequency residuals evaluated over random $(\alpha,\beta)$} &
\textrm{$\sigma$ [Hz]}\\
\hline
fake data (finite calibration length alone): $\delta_{\sigma^+}$ (from \eqref{eq:resFromMPEAlone})&3.4\\
fake data: $\delta_{\sigma^+,G}$ (\eqref{eq:resFromNoieAndMPE})&5\\
measured: $(\delta_m^{m^{\prime}})_s$ (\eqref{eq:rabRes})&13\\
\hline
\end{tabular}
\end{center}
\end{table}

\section{Estimated vector sensitivity from detection noise}
\begin{figure}[htbp]
\centering
\includegraphics[width=1\linewidth]{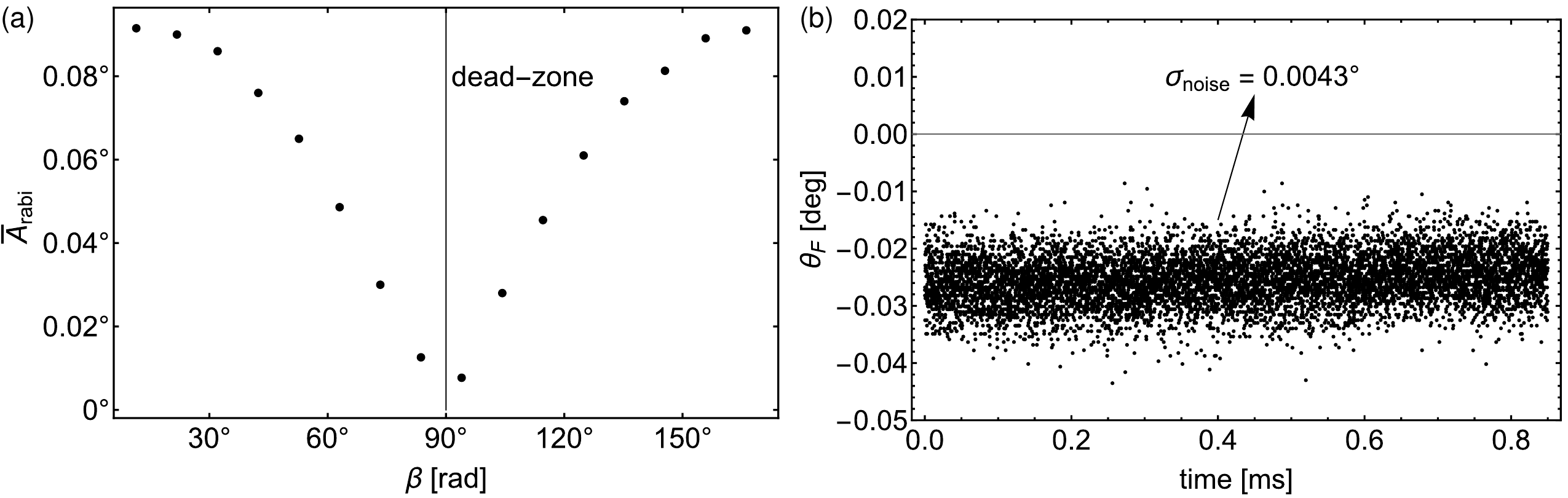}
\caption{Rabi noise characterization (a) The mean Rabi amplitude averaged across all four hyperfine transitions for different polar angles $\beta$. (b) The measured Faraday rotation noise of the polarimeter.}
\label{fig:aveRabAmp}
\end{figure}

In this section, we estimate the contribution of optical detector noise to the vector sensitivity. To accomplish this without unnecessary complexity, we calculate the average Rabi amplitude $\overline{A}_{\text{rabi}}$, which corresponds to an average signal-to-noise ratio (SNR), across the four hyperfine transitions and each magnetic field direction [Fig.~\ref{fig:aveRabAmp}(a)], and approximate all Rabi measurements as having this SNR. The Rabi amplitude is defined as the fitted sinusoidal amplitude of the measured Rabi oscillation. We find that the average Rabi signal amplitude is $\langle \overline{A}_{\text{rabi}} \rangle = 0.061 \degree$. 

Our detector noise is characterized by a standard deviation $\sigma_{\theta}=0.0043\degree$ measured over the duration of a Rabi oscillation measurement [Fig.~\ref{fig:aveRabAmp}(b)]. This detector noise is a bit higher than the shot-noise limit characterized by the sensitivity~\cite{lucivero2014shot} 
\begin{equation}
\label{eq:shotnoise}
S_{\theta_F,\text{shot}}=\frac{1}{2}\frac{1}{\sqrt{\dot{N}_{ph}}}
\end{equation}
where $\dot{N}_{ph}=P/(hc/\lambda)$ is the number of photons in the probe beam per second, $P=1$ mW is the approximate total optical power falling on the polarimeter. With these parameters, \eqref{eq:shotnoise} predicts the shot noise contribution to the Faraday rotation signal to have a standard deviation $\sigma_{shot}=S_{\theta_{F},\text{shot}}\sqrt{f_s/2}=0.001\degree$, where $f_s=10$ MHz is the polarimeter sampling rate.

Next, we determine the Cram\'{e}r-Rao lower-bound (CRLB) variance $\sigma_{\Omega}^2$ to estimate the the frequency uncertainty of a Rabi oscillation. The CRLB for a damped sinusoid is given by~\cite{gemmel2010ultra,grujic2015sensitive} 
\begin{equation}
\label{eq:CRLB}
\sigma_{\Omega}^2\geq \frac{12}{(2\pi)^2 (A_{\text{rabi}}/\sigma_{\theta})^2 f_s T_r^3}C
\end{equation}
where $A_{\text{rabi}}/\sigma_{\theta}$ is the signal-to-noise ratio (SNR), $T_r=0.85$ ms is the measurement time, and $f_s=10$ MHz is the sampling rate. The factor $C$ is an overall constant given in terms of the dephasing time $\gamma_2=1/T_2$ and the number of samples $N=f_s T_r=8500$:
\begin{equation}
C=\frac{N^3}{12}\frac{(1-z^3)^3(1-z^{2N})}{z^2(1-z^{2N})^2-N^2z^{2N}(1-z^2)^2}
\end{equation}
with $z=e^{-\gamma_2/f_s}$. With \eqref{eq:CRLB}, we estimate that the correspond Rabi frequency noise to be $\sigma_{\Omega}=4.3$ Hz for our measured detector noise and $\sigma_{\Omega}=1.0$ Hz for the shot noise limit.

\begin{figure}[!htbp]
\centering
\includegraphics[width=.6\linewidth]{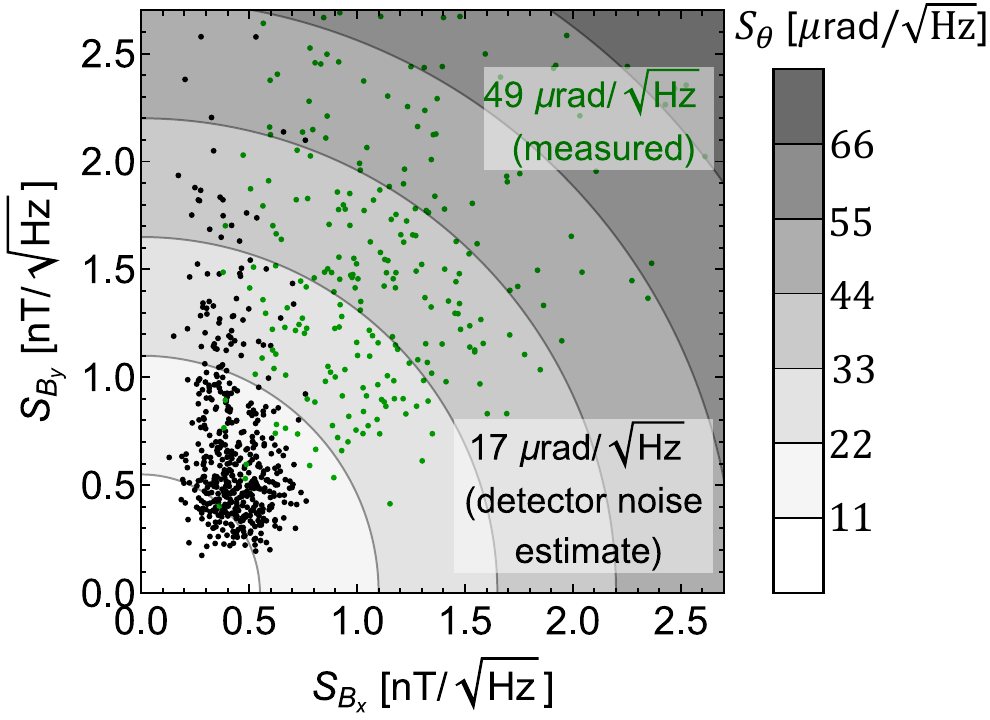}
\caption{Vector component sensitivity estimated from detector noise (black) compared to measured vector sensitivity (green) over the same magnetic field orientations. The text labels denote the average directional sensitivities.}
\label{fig:noiseEstFig}
\end{figure}

Next, we use mean calibrated MPE parameters shown Table~\ref{tab:MPE} to generate a data set of fake Rabi rates $\Omega_m^{m^{\prime}}$ for each hyperfine transition with random noise $\sigma_{\Omega}$ using 
\begin{equation}
(\Omega_m^{m^{\prime}}) =  G \Big( \mu_m^{m^{\prime}} |\mathcal{B}_{k}^{(\alpha,\beta)}|/h ,(\sigma_{\Omega})^2 \Big).
\end{equation}
Here $G(\mu,\sigma^2)$ denotes a normal distribution with mean $\mu$ and variance $\sigma^2$. To avoid further complication in this analysis, we work with Rabi rates instead of generalized Rabi frequencies. From this fake data set of noisy Rabi rates, we solve for $(\alpha,\beta)$ from the system of 12 equations (three MPE configurations and four hyperfine transitions) using the model $\Omega_{m}^{m^{\prime}} = \mu_m^{m^{\prime}} |\mathcal{B}_{k}^{(\alpha,\beta)}|/h$.

The fluctuations in $(\alpha,\beta)$ are converted into vector component fluctuations that are characterized by a standard deviation $\sigma_{B_i}$. From the measurement times $t_m=150$ ms (including the FID measurement time), we calculate the vector component sensitivities through $S_{B_i}=\sigma_{B_i}\sqrt{2t_m}$. These simulated vector component sensitivities are plotted in Fig.~\ref{fig:noiseEstFig} along with the measured vector sensitivities corresponding to the same magnetic field orientations. From Fig.~\ref{fig:noiseEstFig} we find that the mean directional sensitivity is 17 $\mu\text{rad}/\sqrt{\text{Hz}}$, averaged across all DC field directions, due solely to detector noise. This is about three times lower than the mean measured directional sensitivity of 49 $\mu\text{rad}/\sqrt{\text{Hz}}$. This discrepancy is likely due to technical microwave noise. By taking the ratio of the measured detector noise ($0.0043^{\circ}$) to the photon shot noise limit ($0.001^{\circ}$), we deduce a shot noise-limited mean directional sensitivity of 3.9 $\mu\text{rad}/\sqrt{\text{Hz}}$.

\begin{table}[h]
\caption{\label{tab:MPE}
Initial calibrated MPEs used to estimate vector sensitivity from detector noise.}
\begin{center}
\begin{tabular}{|c||c|c|c|c|c|}
\hline
\textrm{MPE} &
\textrm{$\mathcal{B}_x$ $[\mu T]$}&
\textrm{$\mathcal{B}_y$ $[\mu T]$}&
\textrm{$\mathcal{B}_z$ $[\mu T]$}&
\textrm{$\phi_x$ $[\text{rad}]$}&
\textrm{$\phi_y$ $[\text{rad}]$}\\
\hline
1&2.3222& 3.9520& 0.0601& 2.6210& 4.010\\
2&3.5027& 2.3923& 0.0560& 3.0864& 1.7690\\
3&5.5730& 1.6342& 0.0994& 2.7799& 4.7449\\
\hline
\end{tabular}
\end{center}
\end{table}

\section{Possible systematic error sources}
\label{sec:systematicSources}
Some systematic errors in the vector magnetometry data are expected due to the uncertianty in the microwave parameters from the finite MPE calibration length investigated in Sec.~\ref{sec:rabiModelErrors} of this Supplement. From the analysis in Sec.~\ref{sec:rabiModelErrors}, residuals between generalized Rabi frequency measurements and the eigenvalue model $\delta \lambda_m^{m^{\prime}}$ are predicted to have a standard deviation of $\sigma=5$ Hz, which is lower than the measured fluctuations of $\sigma=13$ Hz reported in Fig.~3(c) of the main text. Furthermore, these $\sigma=13$ Hz residuals are fairly consistent with the mean vector magnetometry accuracy of $0.46$ mrad reported in Sec.~7 of the main text, which roughly correlates to $50\text{ kHz}\cdot0.00046\approx 23$ Hz systematic errors for a $50$ kHz Rabi frequency measurement. In the subsections below we highlight potential sources of systematic errors in the Rabi oscillation measurements. We tabulate approximate sizes of these various sources  in Table~\ref{tab:systematicErrors}. The approximate estimate of 15 Hz for the total contribution from all of these systematic errors are in agreement with the back-of-the-envelope 23 Hz error.

\begin{table}[h]
\caption{\label{tab:systematicErrors}
A list of potential systematic error sources and their estimated magnitudes for a 50 kHz Rabi frequency. Details behind these estimates are found in the cited subsections.}
\begin{center}
\begin{tabular}{|c||c|}
\hline
\textrm{Source} &
\textrm{Systematic error [Hz]}\\
\hline
Finite MPE calibration length (Sec.~\ref{sec:rabiModelErrors}) &3.4\\
Microwave phase noise (Sec.~7\ref{sec:micPhaseNoise})&1\\
DC magnetic field inhomogeneity (Sec.~7\ref{sec:magInhomo})&1\\
Microwave spatial inhomogeneity (Sec.~7\ref{sec:micInhomo})&8\\
Rabi frequency fitting errors (Sec.~7\ref{sec:rabiLineshapeErr}) & 2\\
Beyond rotating wave effects (Sec.~7\ref{sec:beyondRWA})  & 0.1\\
\hline
\textbf{Total} & $\approx15$ \\
\hline
\end{tabular}
\end{center}
\end{table}

\subsection{Microwave Phase Noise}
\label{sec:micPhaseNoise}
Noise in our microwave drive is characterized by amplitude and phase noise. In a two-level system this noise affects the Hamiltonian describing Rabi oscillations with Rabi rate $\Omega$ and detuning $\Delta$ by
\begin{equation}
\label{eq:phaseNoiseHamiltonian}
H=\frac{h}{2}(\Omega+\delta\Omega(t))\sigma_x+\frac{h}{2}(\Delta+\delta \Delta(t))\sigma_z
\end{equation}
where $\sigma_i$ are the Pauli spin matrices. The effect of amplitude noise leads to Rabi rate fluctuations $\delta \Omega(t)$ and phase noise leads to detuning fluctuations $\delta \Delta (t)$. Detuning fluctuations might lead to nuanced behavior that causes a shift in the generalized Rabi frequency ($\tilde{\Omega}$). In this section we investigate this effect for the phase noise of our microwave source. Note that in this section, we change the notation so that $\Omega$ and $\Delta$ represent angular frequencies, whereas in the rest of this Supplemental material and in the main text, $\Omega$ and $\Delta$ denote real frequencies in units of Hertz.

\begin{figure}[htbp]
\centering
\includegraphics[width=1\linewidth]{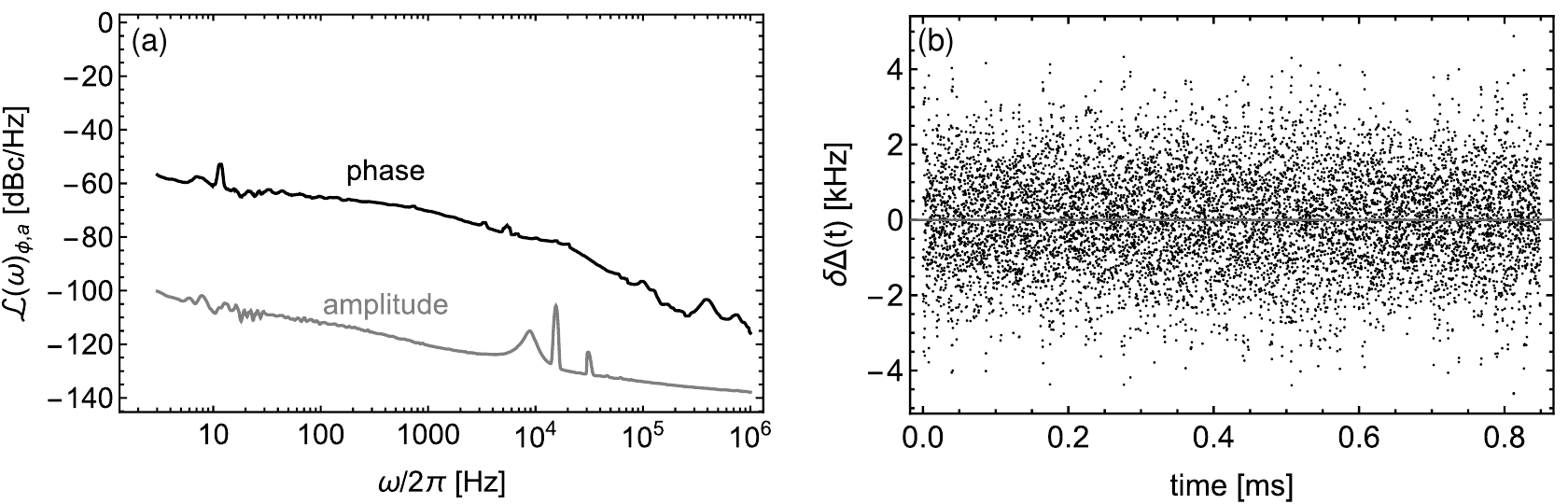}
\caption{Microwave phase noise characterization. (a) The single-sideband phase and amplitude noise of our microwave source. (b) Microwave detuning fluctuations sampled at 10 MHz that are deduced from the measured phase phase in (a). The black line a moving average of 500 points.}
\label{fig:phaseNoise}
\end{figure}

For our microwave source (WindFreak SynthHD Pro + Mini-Circuits ZVE-8G+), these noise sources are characterized by the single-sideband (SSB) phase $\mathcal{L}_{\phi}$ and amplitude $\mathcal{L}_{a}$ noise displayed in Fig.~\ref{fig:phaseNoise}(a) measured with a Rohde \& Schwarz FSWP26 phase noise analyzer at 6835.82 MHz. In the following analysis, we focus solely on the effects of $\mathcal{L}_{\phi}$, as $\mathcal{L}_{a}$ is not expected to contribute to systematic errors in the Rabi measurements and is significantly smaller than phase noise overall. We note that the WindFreak SynthHD Pro phase noise measured in Fig.~\ref{fig:phaseNoise}(a) is much higher than its specifications. We found that this descrepancy is due to additional noise in a malfunctioning 10 MHz atomic clock reference.

As shown in Ref.~\cite{ball2016role}, the SSB phase noise is related to the unilateral power spectral density (PSD) of the corresponding detuning fluctuations $\delta \Delta (t)$ by $S_{\Delta}(\omega)=\frac{\omega^2}{2}10^{\frac{\mathcal{L}_{\phi}(\omega)}{10}}$. This PSD can be converted into a time-series by reflecting $S_{\Delta}(\omega)$ to the negative $-\omega$ frequencies, and creating a complex noise spectral density $N(\omega)=e^{i\phi_r}\sqrt{S_{\Delta}(\omega)/2}$, where $\phi_r\in[0,2\pi]$ is a random phase evaluated at each $\omega$. These random phases are chosen such that $N(\omega)=N(-\omega)^*$. By taking the inverse Fourier transform of $N(\omega)$ we obtain a time-series for $\delta \Delta(t)$ [Fig~\ref{fig:phaseNoise}](b). 

\begin{figure}[!htbp]
\centering
\includegraphics[width=.5\linewidth]{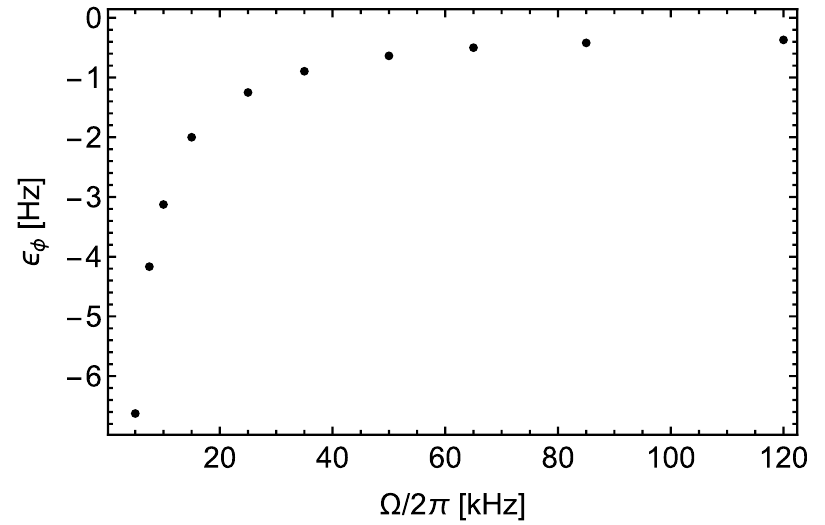}
\caption{A plot of the calculated shift in the generalized Rabi frequency, driven on-resonance, due to the measured microwave phase noise shown in Fig.~\ref{fig:phaseNoise}a.}
\label{fig:phaseNoiseError}
\end{figure}

Next, we simulate Rabi oscillations driven by the two-level Hamiltonian in \eqref{eq:phaseNoiseHamiltonian} with the following density matrix equations
\begin{align}
\begin{split}
\label{eq:denMatTwo}
\dot{\rho}_{11}&=i\frac{\Omega}{2}(\rho_{12}-\rho_{21})+\Gamma \rho_{22}\\
\dot{\rho}_{12}&=i\frac{\Omega}{2}(\rho_{11}-\rho_{22})-(\Gamma/2+i\delta\Delta(t)) \rho_{12}\\
\dot{\rho}_{21}&=i\frac{\Omega}{2}(\rho_{22}-\rho_{11})-(\Gamma/2-i\delta\Delta(t)) \rho_{21}\\
\dot{\rho}_{22}&=-i\frac{\Omega}{2}(\rho_{12}-\rho_{21})-\Gamma \rho_{22}
\end{split}
\end{align}
where we set $\delta \Omega (t)=0$, $\Delta=0$, and $\Gamma/2\pi =1000$ $ s^{-1}$. The density matrix is initialized at $t=0$ to $\rho_{22}=1$ and $\rho_{11}=\rho_{12}=\rho_{21}=0$. We numerically solve \eqref{eq:denMatTwo} over 0.85 ms and update $\delta \Delta(t)$ according to phase noise fluctuations shown in Fig.~\ref{fig:phaseNoise}(b) at time-intervals of $\delta t =1/f_s$ defined by our experimental sample rate $f_s=10$ MHz. Fig.~\ref{fig:phaseNoiseError} plots the error in the fitted generalized Rabi frequency of the simulated Rabi oscillation for different Rabi rates $\Omega$. From Fig.~\ref{fig:phaseNoiseError} we take 1 Hz as the systematic error from microwave phase noise.

\subsection{Magnetic field spatial inhomogeneity}
\label{sec:magInhomo}
Magnetic field spatial inhomogeneity will cause the detuning ($\Delta(x,y,z)$) of the microwave drive to vary across the vapor cell volume. Thus, the Rabi oscillation measurement is an average of independent Rabi oscillations with various detunings. Because the generalized Rabi frequency ($\tilde{\Omega}$) only increases with detuning, $\tilde{\Omega}$ will be positively shifted when the microwave detuning averaged over the vapor cell volume is zero. This shift will appear as a systematic error.

We characterize the spatial inhomogeneity of the magnetic field within our coil system with the normalized magnetic field gradient defined as
\begin{equation}
G=\frac{1}{B_i}\sqrt{\bigg(\frac{\partial B_i}{\partial x}\bigg)^2+\bigg(\frac{\partial B_i}{\partial y}\bigg)^2+\bigg(\frac{\partial B_i}{\partial z}\bigg)^2}.
\label{eq:gValues}
\end{equation}
Maximum and minimum values of $G$ for different coil pair excitations are listed in Table~\ref{tab:coilGradient} with the last column showing the estimated magnetic field variation from $(G_{\text{min}}+G_{\text{max}})/2\times l_i$ over the $l_{x,y}=3$ mm and $l_z=2$ mm dimensions of the microfabricated cell. Measurements of the magnetic field spatial homogeneity were performed with a fluxgate vector magnetometer mounted to a 3-axis translation stage. 

\begin{table}[!h]
\caption{\label{tab:coilGradient}
Table of maximum and minimum G-values defined in Eq.~\ref{eq:gValues}. The last column shows the estimated magnetic field variation from $(G_{\text{min}}+G_{\text{max}})/2\times l_i$ over the $l_{x,y}=3$ mm and $l_z=2$ mm dimensions of the cell.}
\begin{center}
\begin{tabular}{ | c | c| c | c|} 
  \hline
  coil& $G_{\text{min}}$ (mm$^{-1}$) & $G_{\text{max}}$ (mm$^{-1}$)& $\delta B_i$ [nT] \\
  \hline
  x saddle& 0.005\% & 0.036\% & 31 \\ 
  \hline
  y saddle & 0.001\% & 0.017\%& 14 \\ 
  \hline
  z Helmholtz & 0.047\% & 0.055\% & 51\\ 
  \hline
\end{tabular}
\end{center}
\end{table}
 
Considering the field variation in Table~\ref{tab:coilGradient} of 51 nT across the 2 mm cell length, this would result in a detuning variation of approximately 1 kHz for the $\ket{1,1}-\ket{2,2}$ transition, calculated as $(51\text{ nT}) \times (7\text{ kHz/nT}) \times 3 \approx 1$ kHz, assuming $\vec{B}_{\text{dc}} \parallel \hat{z}$. For a 50 kHz Rabi frequency driven near-resonance this would cause an approximate systematic shift of $50-\sqrt{50^2+0.25^2}=0.0006$ kHz assuming the two-level Rabi formula (Eq.~2 in the main text). Therefore, we conclude that systematic errors from DC magnetic field spatial gradients are at the 1 Hz level.
\subsection{Microwave inhomogeneity}
\label{sec:micInhomo}
While the Rabi rate defined in the main text
\begin{equation}
\label{eq:rabiRateComplex}
\Omega_m^{m^{\prime}}=\frac{\mu_m^{m^{\prime}}}{h}\mathcal{B}_k=\frac{\mu_m^{m^{\prime}}}{h}R_y(-\beta)R_z(-\alpha)\vec{\mathcal{B}}\cdot\epsilon_k
\end{equation}
is a complex number, the Rabi oscillation frequencies measured from the Faraday signal are real numbers. This is akin to measuring $|\Omega_m^{m^{\prime}}|$. For two complex numbers $c_1$ and $c_2$, it is generally true that $|c_1|+|c_2|\neq|c_1+c_2|$. Similarly if there is microwave inhomogeneity, this results in a spatially varying Rabi frequency $\Omega_m^{m^{\prime}}(x,y,z)$, such that the $|\Omega_m^{m^{\prime}}|$ that we measure, neglecting the probe beam profile, is
\begin{equation}
|\Omega_m^{m^{\prime}}|=\frac{1}{V}\int_V |\Omega_m^{m^{\prime}}(x,y,z)| dV\neq \frac{1}{V}\Big|\int_V \Omega_m^{m^{\prime}}(x,y,z) dV\Big|.
\end{equation}
Thus, there is no single microwave phasor $\vec{\mathcal{B}}$ that accurately models the directional dependence of $|\Omega_m^{m^{\prime}}|$ within \eqref{eq:rabiRateComplex} when there is microwave spatial inhomogeneity.
\begin{figure}[htbp]
\centering
\includegraphics[width=1\linewidth]{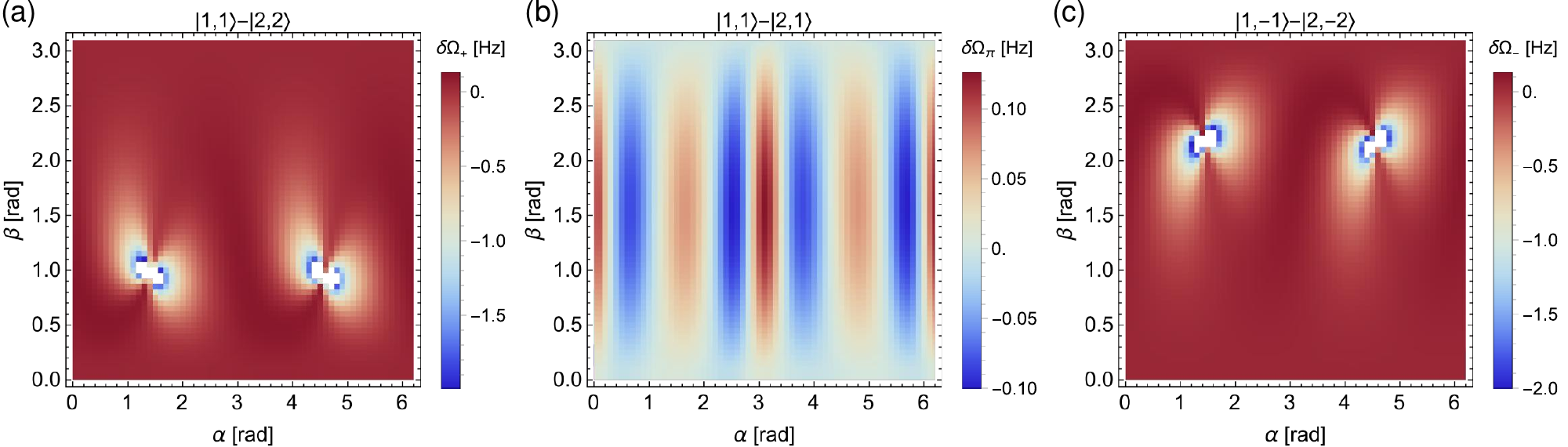}
\caption{Predicted systematics of the Rabi frequencies assuming microwave amplitude ($\mathcal{B}_{x}$,$\mathcal{B}_{y}$,$\mathcal{B}_{z}$) inhomogeneities for the (a) $\sigma^+$, (b) $\pi$, and (c) $\sigma^-$ hyperfine transitions.}
\label{fig:inhomAmp}
\end{figure}

\begin{figure}[htbp]
\centering
\includegraphics[width=1\linewidth]{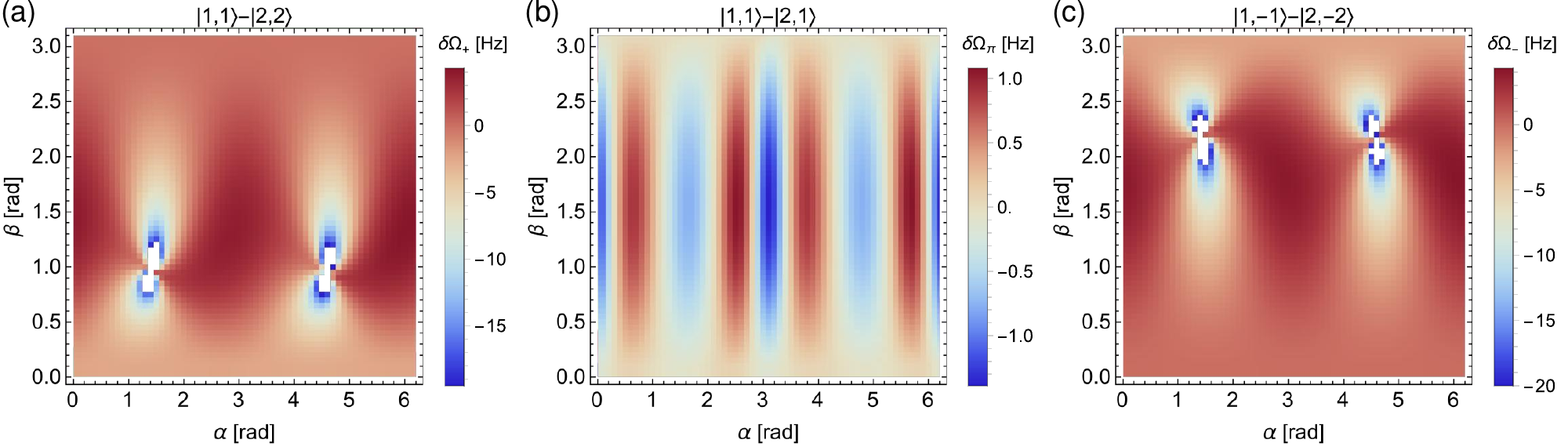}
\caption{Predicted systematics of the Rabi frequencies assuming microwave phase ($\phi_{x}$,$\phi_{y}$) inhomogeneities for the (a) $\sigma^+$, (b) $\pi$, and (c) $\sigma^-$ hyperfine transitions.}
\label{fig:inhomPhase}
\end{figure}

We test this by generating spatially dependent MPE 1 parameters from Table~\ref{tab:MPE} assuming that the spatial variation is only in the z-coordinate for simplicity. These variations are expressed to second-order
\begin{align}
B_i(x)=B_{i,0}(1+a_1 z+a_2 z^2)\\
\phi_i(x)=\phi_{i,0}(1+b_1 z+b_2 z^2)
\end{align}
where $i=x,y,z$ and $a_1=1$ nT/mm, $a_2=1$ nT/mm$^2$, $b_1=1$ mrad/mm, and $b_2=2$ mrad/mm$^2$. The actual spatial inhomogeneity is difficult to characterized so these numbers are very rough guesses. Then over all field directions $(\alpha,\beta)$ we calculate the averaged spherical microwave component $\overline{\mathcal{B}}_k$ over the $2$ mm z-dimension of the MEMS cell
\begin{equation}
\overline{\mathcal{B}}_k=\frac{1}{2}\int_{-1}^1\mathcal{B}_k(z)dz
\end{equation}
where $k=\sigma^{\pm},\pi$. These averaged spherical  microwave components are then converted to real Rabi rates through the absolute value of \eqref{eq:rabiRateComplex}. 

Next, we fit MPE parameters for each $k$ to the resulting $\overline{\mathcal{B}}_k$ using the real Rabi rate $|\Omega_m^{m^{\prime}}|$ predicted from \eqref{eq:rabiRateComplex}. The discrepancy between the averaged $|\Omega_m^{m^{\prime}}|$ calculated with \eqref{eq:rabiRateComplex}, and that predicted with the MPE parameter fits of the spatially-averaged data is shown in Fig.~\ref{fig:inhomAmp} for only microwave amplitude variations ($b_{1,2}=0$) and Fig.~\ref{fig:inhomPhase} for only phase variations ($a_{1,2}=0$). With these parameters, we see the largest systematic errors occur with the phase variations at the 5-10 Hz scale. Not considered here is how much the microwave inhomogeneity also modifies the exponential decay of the Rabi oscillation, which could also introduce its own systematic shifts in the Rabi frequency fitting. While it is challenging for us to estimate the exact microwave inhomogeneity in our experiment, this study demonstrates its importance when trying to make accurate vector measurements.

\subsection{Collisional effects on the Rabi lineshape}
\label{sec:rabiLineshapeErr}
The primary dephasing mechanism for Rabi oscillations driven in hot atomic vapor, are Rb-Rb spin-exchange collisions. In addition, wall collisions, spin-destruction collisions arising from Rb-Rb and Rb-N$_2$ collisions, and Carver relaxation due to Rb-N$_2$ collisions also contribute. In our Rabi frequency analysis, we fit Rabi oscillations ($f\rightarrow \tilde{\Omega}$) using the following time-domain decay model
\begin{align}
\begin{split}
\label{eq:timeDomainFit}
\theta_F(t)=&a_0+a_{1}e^{-t/t_{1}}+a_{2}e^{-t/t_{2}}+\\&e^{-t/t_3}(a_3 \text{sin}[2\pi t f]+a_4 \text{cos}[2\pi t f])
\end{split}
\end{align}
Spin-exchange collisions can give rise to multiple dephasing time-constants instead of a single $t_3$ dephasing time in \eqref{eq:timeDomainFit}. This can lead to systematic errors in the time-domain fitting. 
\begin{figure}[htbp]
\centering
\includegraphics[width=.7\linewidth]{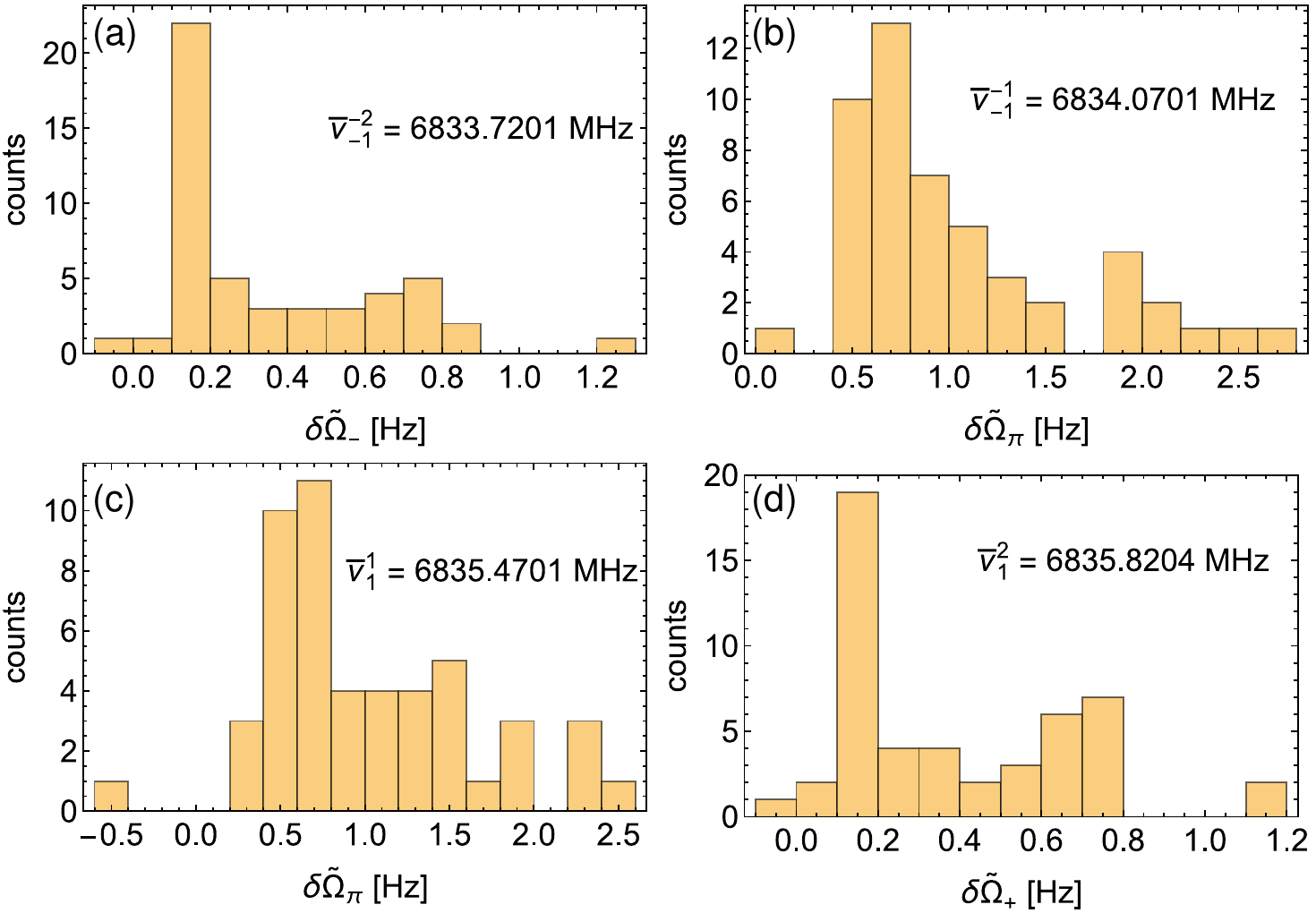}
\caption{Errors between fits of simulated Rabi oscillations with the $\delta\lambda_m^{m^{\prime}}$ for the (a) $\sigma^-$ (b,c) $\pi$ and (d) $\sigma^+$ hyperfine transitions.}
\label{fig:fitAccuracy}
\end{figure}

We model Rabi oscillations with the following Liouville equation that includes decoherence from spin-exchange ($\Gamma_{\text{se}}=3890\text{}s^{-1}$), spin-destructive collisions ($\Gamma_{\text{sd}}=41\text{}s^{-1}$), and wall collisions ($\Gamma_{\text{D}}=450\text{}s^{-1}$)
\begin{align}
\label{eq:timeEvoSupp1}
\begin{split}
    \frac{d\rho}{dt}=&\frac{[H^{(\alpha,\beta)},\rho]}{i \hbar}+\Gamma_{\text{se}}\big(\tilde{\phi}(1+4\langle \tilde{\mathbf{S}} \rangle\cdot \tilde{\textbf{S}})-\rho\big)+\Gamma_{\text{sd}}\big(\tilde{\phi}-\rho\big)+\Gamma_D(\rho^e-\rho)
\end{split}
\end{align}
which is discussed in further detail in \cite{kiehl2023coherence}. The calculation of collision rates $\Gamma_{\text{se,sd,wall}}$ are discussed in Ref.~\cite{kiehl2024correcting}. Here, $\tilde{\mathcal{O}}$ denotes operators made time-independent through the rotating-wave approximation (RWA), which is detailed in the Supplementary Material of Ref.~\cite{kiehl2023coherence}. The rotated atom-microwave Hamiltonian $H^{(\alpha,\beta)}=e^{-i\alpha F_z}e^{-i\beta F_y}H e^{i\beta F_y}e^{i\alpha F_z}$ is defined in terms of hyperfine angular momentum operators defined over the $F=1,2$ manifolds $F_{y,z}$ and the same atom-microwave Hamiltonian $H$ defined in the main text. For these simulations we neglect Carver relaxation due to its small effect predicted for our 180 Torr buffer gas pressure~\cite{kiehl2023coherence}. Here, $\phi=\rho/4+\textbf{S}\cdot \rho\textbf{S}$ is known as the nuclear part of the density matrix and $\rho^e$ is the equilibrium density matrix
with all populations $\rho^e_{ii}$ being equal. For simplicity we assume that the initial atomic state ${\rho_0}$ is the diagonalized density matrix with $F=2$ populations depopulated and the $F=1$ populations equally populated.

We solve \eqref{eq:timeEvoSupp1} by using MPE 1 parameters in \ref{tab:MPE} to evaluate spherical microwave components for each of the field directions considered in the main text. A histogram of the discrepancy between the generalized Rabi frequencies fitted to the simulated Rabi oscillations compared to the $\delta \lambda_m^{m^{\prime}}$ model is shown in Fig.~\ref{fig:fitAccuracy} for each of the four hyperfine transitions.

\subsection{Beyond RWA systematics}
\label{sec:beyondRWA}
Another potential systematic, though likely having a smaller effect than the above systematic list are beyond rotating wave effects. An rough estimate of the scale of shift from beyond RWA is given by the Bloch-Siegert shift
\begin{equation}
\frac{1}{4}\frac{(\Omega_m^{m^{\prime}})^2}{\nu_m^{m^{\prime}}}\approx 0.1 \text{ Hz} 
\end{equation}
assuming $\Omega_m^{m^{\prime}}\approx 50$ kHz and $\nu_m^{m^{\prime}}\approx6.834$ GHz.

\section{Unique determination of the magnetic field direction with MPE\MakeLowercase{s}}
\begin{figure}[htbp]
\centering
\includegraphics[width=1\linewidth]{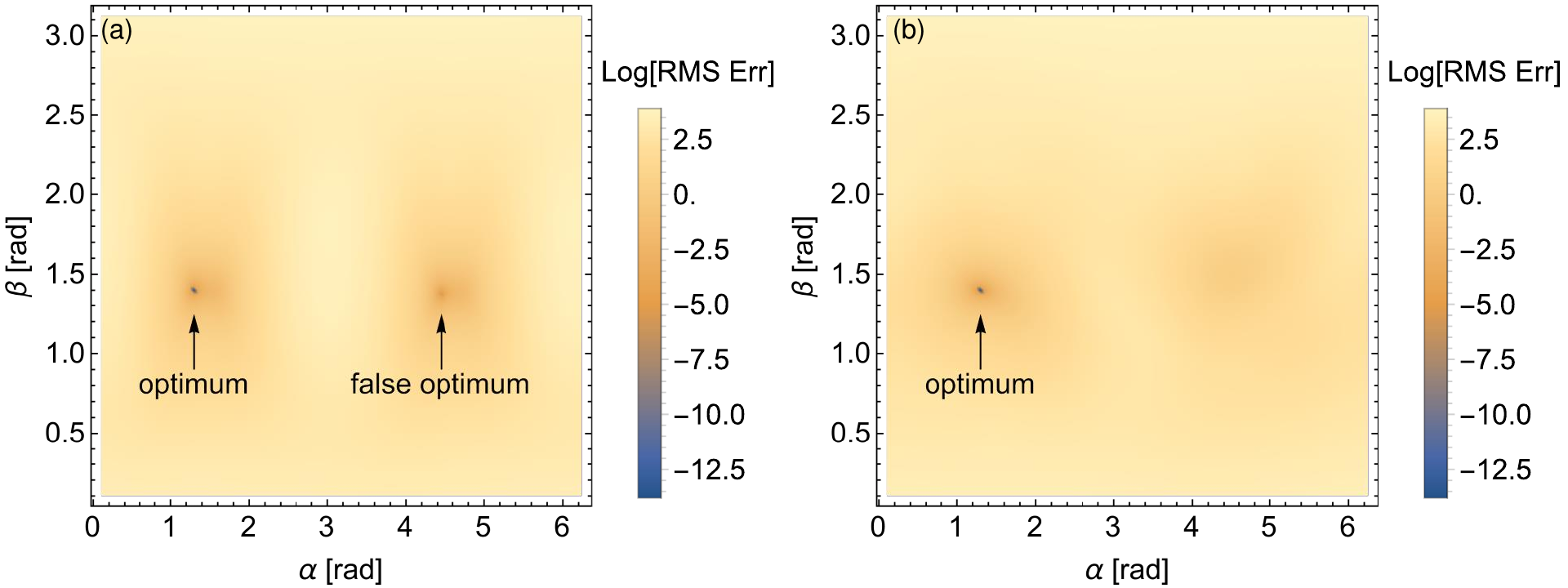}
\caption{Unique vector detection with Rabi measurements. Evaluation of the cost function \eqref{eq:vectorCost} for (a) the MPEs in Table~\ref{tab:MPE} and (b) replacing MPE 3 parameter with an MPE that contains a significant $\mathcal{B}_z$ component. In both of these evaluations we observe a unique global minimum.}
\label{fig:optimumSol}
\end{figure}
We evaluate the uniqueness of the $(\alpha,\beta)$ fitted from Rabi measurements by evaluating the following cost function
 \begin{align}
 \label{eq:vectorCost}
\begin{split}
&\sum_{s=1}^3 \sum_{m,m^{\prime}} \Big(|R_y(-\beta)R_z(-\alpha)(\vec{\mathcal{B}}_m^{m^{\prime}})_s\cdot\epsilon_k\big| -|R_y(-\beta_1)R_z(-\alpha_1)(\vec{\mathcal{B}}_m^{m^{\prime}})_s\cdot\epsilon_k\big| \Big)^2
\end{split}
\end{align}
which compares the spherical microwave components, calculated from the MPE parameters in Table~\ref{tab:MPE}, at $(\alpha_1,\beta_1)=(1.3,1.4)$ rad with all other field directions. Plotting this cost function in Fig.~\ref{fig:optimumSol}(a) reveals a unique global solution at the correct $(\alpha_1,\beta_1)$. There is also a local minimum that would be a false optimum value for the field direction. Instead of using MPEs all in a plane, we found that by replacing MPE 3 in Table~\ref{tab:MPE} with $(\mathcal{B}_x,\mathcal{B}_y,\mathcal{B}_z,\phi_x,\phi_y)=(1\text{ $\mu$T},2\text{ $\mu$T},3\text{ $\mu$T},0\text{ rad},0.5\text{ rad})$ gives a more pronounced optimum solution [Fig.~\ref{fig:optimumSol}(b)].

\section{Expressing MPE 3 in terms of  MPE 1 and MPE 2}
In this work, MPE 1 and MPE 2, with microwave phasors represented as $\vec{\mathcal{B}}_1$ and $\vec{\mathcal{B}}_2$, refer to the microwave fields generated by exciting either of two microwave ports, each designed to couple to an unique orthogonally polarized mode within the cavity's xy-plane. The microwave field for MPE 3 is generated by simultaneously exciting both of these ports. Thus, as discussed in Sec.~8 of the main text, the calibrated MPE 3 parameters, represented by the microwave phasor $\vec{\mathcal{B}}_3$, are well-approximated by summing the microwave fields predicted from the parameters of MPE 1 and MPE 2. This process is mathematically formulated as
\begin{align}
\begin{split}
\vec{\mathcal{B}}_3=&\vec{\mathcal{B}}_1+\vec{\mathcal{B}}_2=\{\mathcal{B}_{x,1}e^{-i\phi_{x,1}},\mathcal{B}_{y,1}e^{-i\phi_{y,1}},\mathcal{B}_{z,1}\}+\\
&\{\mathcal{B}_{x,2}e^{-i(\phi_{x,2}+\phi_r)},\mathcal{B}_{y,2}e^{-i(\phi_{y,2}+\phi_r)},\mathcal{B}_{z,2}e^{-i\phi_{r}}\}.
\end{split}
\end{align}

The MPE 3 parameters are then given by
\begin{align}
\begin{split}
\label{eq:sumPorts}
\mathcal{B}_{x,3}=\sqrt{(\mathcal{B}_{x,1}\text{cos}(\phi_{x,1})+\mathcal{B}_{x,2}\text{cos}(\phi_{x,2}+\phi_{r}))^2+(\mathcal{B}_{x,1}\text{sin}(\phi_{x,1})+\mathcal{B}_{x,2}\text{sin}(\phi_{x,2}+\phi_r))^2}\\
\mathcal{B}_{y,3}=\sqrt{(\mathcal{B}_{y,1}\text{cos}(\phi_{y,1})+\mathcal{B}_{y,2}\text{cos}(\phi_{y,2}+\phi_r))^2+(\mathcal{B}_{y,1}\text{sin}(\phi_{y,1})+\mathcal{B}_{y,2}\text{sin}(\phi_{y,2}+\phi_r))^2}\\
\mathcal{B}_{z,3}=\sqrt{(\mathcal{B}_{z,1}+\mathcal{B}_{z,2}\text{cos}(\phi_r))^2+(\mathcal{B}_{z,2}\text{sin}(\phi_r))^2}\\
\phi_{x,3}=\text{arg}(\mathcal{B}_{x,1}\text{cos}(\phi_{x,1})+\mathcal{B}_{x,2}\text{cos}(\phi_{x,2}+\phi_{r})+i(\mathcal{B}_{x,1}\text{sin}(\phi_{x,1})+\mathcal{B}_{x,2}\text{sin}(\phi_{x,2}+\phi_r)))\\
\phi_{y,3}=\text{arg}(\mathcal{B}_{y,1}\text{cos}(\phi_{y,1})+\mathcal{B}_{y,2}\text{cos}(\phi_{y,2}+\phi_{r})+i(\mathcal{B}_{y,1}\text{sin}(\phi_{y,1})+\mathcal{B}_{y,2}\text{sin}(\phi_{y,2}+\phi_r)))\\
\phi_{z,3}=\text{arg}(\mathcal{B}_{z,1}+\mathcal{B}_{z,2}\text{cos}(\phi_{r})+i(\mathcal{B}_{z,2}\text{sin}(\phi_r)))
\end{split}
\end{align}

To compare to calibrated MPE 3 parameters that assume $\phi_{z,3}=0$ we make the transformation
\begin{align}
\begin{split}
\label{eq:phaseShift}
\phi_{x,3}\rightarrow \phi_{x,3}-\phi_{z,3}\\
\phi_{y,3}\rightarrow \phi_{y,3}-\phi_{z,3}.
\end{split}
\end{align}
For initial calibrated MPE 1 and MPE 2 parameters in Table~\ref{tab:MPE} we calculate with \eqref{eq:sumPorts} and \eqref{eq:phaseShift} $(\mathcal{B}_{x,3},\mathcal{B}_{y,3},\mathcal{B}_{z,3})=(5.5820,1.6327,0.1002)$ $\mu$T,  $(\phi_{x,3},\phi_{y,3})=(2.7707,4.7489)$ rad, and $\phi_r=1.0580$ rad. These values closely match the independently calibrated MPE 3 parameters, as listed in Table~\ref{tab:MPE}.